\documentclass[journal]{IEEEtran}
\usepackage{graphicx} 
\usepackage{xcolor}
\usepackage{acronym}
\usepackage{amsmath}
\usepackage{amssymb}
\usepackage{array}
\usepackage{balance}
\usepackage{pifont}
\usepackage{subcaption}
\usepackage{wasysym} 
\usepackage[hang,flushmargin]{footmisc}

\acrodef{UAV}{Unmanned Aerial Vehicle}
\acrodef{BER}{Bit-Error-Rate}
\acrodef{PDR}{Packet Delivery Rate}
\acrodef{SNR}{Signal to Noise Ratio}
\acrodef{SDR}{Software-Defined Radio}
\acrodef{RSS}{Received Signal Strength}
\acrodef{PHY}{Physical}
\acrodef{BPSK}{Binary Phase-Shift Keying}
\acrodef{OFDM}{Orthogonal Frequency Division Multiplexing}
\acrodef{NN}{Neural Network}
\acrodef{ANN}{Artificial \ac{NN}}
\acrodef{CNN}{Convolutional \acp{NN}}
\acrodef{AI}{Artificial Intelligence}
\acrodef{ML}{Machine Learning}
\acrodef{DL}{Deep Learning}
\acrodef{RJP}{Relative Jamming Power}
\acrodef{ROR}{Receiver Oversampling Ratio}
\acrodef{JOR}{Jamming Oversampling Ratio}
\acrodef{mse}{mean-squared-error}
\acrodef{GCS}{Ground Control Station}
\acrodef{RPAS}{Remotely-Piloted Aircraft System}
\acrodef{AWGN}{Additive White Gaussian Noise}
\acrodef{AUC}{Area Under the Curve}
\acrodef{ROC}{Receiver Operating Characteristic}
\acrodef{TPR}{True Positive Ratio}
\acrodef{TNR}{True Negative Ratio}
\acrodef{AGC}{Automatic Gain Control}
\acrodef{WSN}{Wireless Sensor Networks}
\acrodef{IoT}{Internet of Things}
\acrodef{CPU}{Central Processing Unit}

\newcommand{\sol}{\emph{BloodHound$+$}}

\newcolumntype{P}[1]{>{\centering\arraybackslash}p{#1}}

\begin{document}
\bstctlcite{IEEEexample:BSTcontrol}

\title{Jamming Detection in Low-BER\\ Mobile Indoor Scenarios via Deep Learning}
\author{
    \IEEEauthorblockN{Savio Sciancalepore\IEEEauthorrefmark{1}\IEEEauthorrefmark{2}, Fabrice Kusters\IEEEauthorrefmark{1}, Nada Khaled Abdelhadi\IEEEauthorrefmark{1}, Gabriele Oligeri\IEEEauthorrefmark{3}}
    \thanks{\noindent\IEEEauthorblockA{\IEEEauthorrefmark{1}Eindhoven University of Technology, Eindhoven, Netherlands. \\
    \IEEEauthorblockA{\IEEEauthorrefmark{2}Eindhoven Artificial Intelligence Systems Institute (EAISI)}. Email: s.sciancalepore@tue.nl, \{f.e.kusters, n.k.s.abdelhadi\}@student.tue.nl}\\
    \IEEEauthorblockA{\IEEEauthorrefmark{3}Division of Information and Computing Technology (ICT), College of Science and Engineering (CSE), Hamad Bin Khalifa University (HBKU), Doha, Qatar. Email: goligeri@hbku.edu.qa}}
}

\IEEEtitleabstractindextext{
\begin{abstract}
    The current state of the art on jamming detection relies on link-layer metrics. A few examples are the bit-error rate (BER), the packet delivery ratio, the throughput, and the signal-to-noise ratio (SNR). As a result, these techniques can only detect jamming \emph{ex-post}, i.e., once the attack has already taken down the communication link. These solutions are unfit for mobile devices, e.g., drones, which might lose the connection to the remote controller, being unable to predict the attack.    
    Our solution is rooted in the idea that a drone unknowingly flying toward a jammed area is experiencing an increasing effect of the jamming, e.g., in terms of BER and SNR. \textcolor{black}{Therefore, drones might use the abovementioned phenomenon to detect jamming before the increase of the BER and the decrease of the SNR completely disrupt the communication link. Such an approach would allow drones and their pilots to make informed decisions and maintain complete control of navigation, enhancing security and safety.}
    This paper proposes Bloodhound$+$, a solution for jamming detection on mobile devices in low-BER regimes. Our approach analyzes raw physical-layer information (I-Q samples) acquired from the wireless channel. We assemble this information into grayscale images and use sparse autoencoders to detect image anomalies caused by jamming attacks. To test our solution against a broad set of configurations, we acquired a large dataset of indoor measurements using multiple hardware, jamming strategies, and communication parameters. Our results indicate that Bloodhound$+$ can detect indoor jamming up to 20 meters from the jamming source at the minimum available relative jamming power, with a minimum accuracy of 99.7\%. Our solution is also robust to various sampling rates adopted by the jammer and to the type of signal used for jamming.
\end{abstract}
}

\maketitle

\IEEEdisplaynontitleabstractindextext


\section{Introduction}
\label{sec:intro}

\textcolor{black}{Drones, a.k.a. \acp{UAV}, represent an evolution of the \ac{IoT} paradigm. Since we define the \ac{IoT} as a network of physical objects (things) that are embedded with sensors, software, and other technologies to connect and exchange data with other devices and systems over the Internet~\cite{iot_ibm},\cite{iot_oracle}, it is immediate to include drones in the IoT domain. They are battery-powered devices, featuring limited onboard storage space and integrating \acp{CPU} with heterogeneous computational capabilities. Recent papers even coined the notion of \emph{Flying IoT} to specifically identify drones as a specification of the IoT paradigm~\cite{genc2017_micro},~\cite{wisse2023_iotj},~\cite{yazdinejad2020_iotj},~\cite{hoque2021_iotj}.
}

\textcolor{black}{Today, drones are increasingly used both outdoors, e.g., for disaster management, search and rescue, and goods delivery~\cite{abualigah2021_sensors}, and indoors, e.g., for inventory management, intra-logistics of items, inspection and surveillance~\cite{wawrla2019_eth}, and leading companies such as IKEA and Amazon are actively experimenting products for warehouse management and home surveillance~\cite{ikea_drones},~\cite{amazon_indoorDrone}. In this context, leading market analysis companies estimate a current market value of up to 100.37 billion USD by 2029, with a compound annual growth rate of 25.5\% in the next years~\cite{market}. }

As their role becomes more central, drones are increasingly the main target of several cybersecurity attacks. In particular, due to the reliance on wireless channels for communication, video streaming, and telemetry, attackers can easily disrupt drones' operation through \emph{jamming} attacks~\cite{multerer2017_eurad}. Jamming can significantly disrupt wireless communications in a given area by injecting high-power noise into the same channel used by legitimate communication parties~\cite{pirayesh2022_comst}. Depending on the firmware onboard, the drone might return to the mission's starting point, land, or even crash, with potential hazards for people in the area, especially for indoor applications~\cite{ferreira2022_sensors}.

\textcolor{black}{At the time of writing, most of the available solutions for jamming detection work by identifying the deterioration of the \ac{BER}, the \ac{PDR}, or the \ac{SNR} of the communication link.} Some of them also use the \emph{spectrogram} of the signal in the \ac{PHY} layer, looking for sudden anomalies (see Section~\ref{sec:related} for an overview). Such solutions work reliably and effectively when applied to static deployments. However, they can detect jamming mainly \emph{ex-post}, i.e., once jamming has already disrupted the regular operations of the communication link. Applying such approaches to drones would trigger the default actions listed above, possibly causing hazards to the drone and surrounding people. 
In this context, mobile devices such as drones typically experience an increasing effect of jamming (increasingly high \ac{BER}, low \ac{PDR}, low \ac{SNR}) while approaching the jammed area. Drones might exploit the mentioned phenomena to deploy a solution to detect jamming in a low-BER regime, i.e., before entering an area where the high-power noise injected by the jammer completely disrupts the communication link. Such a solution is especially critical for indoor applications, where people and drones often work together in limited space.

{\bf Contribution.} In this paper, we propose \sol, an innovative solution for jamming detection in low-BER mobile indoor scenarios leveraging state-of-the-art \ac{DL} techniques. \sol\ allows to carry out jamming detection by converting raw \ac{PHY} data, that is, I-Q samples, into images while detecting anomalies in their shape by resorting to autoencoders. 

\textcolor{black}{Although our approach can apply to any mobile device, we specifically consider the case of drones operating indoors, e.g., for warehouse inspections~\cite{apMoller_indoorDrones}, inspired by the recent deployment of real-world use cases, such as the one managed by IKEA~\cite{ikea_drones} and Amazon~\cite{amazon_indoorDrone}.}
When applied to autonomous or remotely piloted vehicles and drones, \sol\ can detect the approach of a jammed area with very low BER values, allowing the remote entity to detect jamming while maintaining complete control of the communication link. \textcolor{black}{To verify the effectiveness of our proposed approach, we conducted an extensive measurement campaign emulating \acp{UAV} through Ettus Research X310 and LimeSDR radios, using multiple hardware devices (multiple Ettus Research X310 and LimeSDR radios), communication link configurations, and jamming conditions. Using such measurements, we tested the effectiveness of \sol\ and other competing approaches for detecting jamming in a low-BER regime. Our results show that \sol\ can detect jamming in scenarios with a lower BER ($\approx1e-6$) compared to benchmark solutions, e.g., with an accuracy of $0.997$ when the adversary jams at a distance of $10$~m from the target with a \ac{RJP} of $0.1$. 
Our solution is also very robust to: (i) the distance from the jammer, (ii) the training set size, (iii) the number of acquired samples, (iv) the sampling ratios at the jammer and the receiver, (v) the type of jamming signal (tone, Gaussian, or deceptive), as well as (vi) the adoption of different jamming hardware and radio types. We envision that \sol\ can be used during regular operations of the UAV by acquiring physical layer information from the (already existing) UAV communications channel. In addition, it can be activated when desired, so as not to cause significant overhead on the UAV.}

This contribution extends and completes our previous work published in~\cite{alhazbi2023_ccnc} by providing the following new content.
\begin{itemize}
    \item We focus on an indoor scenario, providing a brand new range of data considering additional hardware, modulation techniques, and jamming strategies.
    \item We consider a stronger adversary model, assuming that the adversary knows the sampling rate and modulation techniques of the legitimate communication link. This knowledge allows the adversary to optimize the parameters of the jamming attack to boost its effectiveness and avoid detection simultaneously.
    \item We design a new optimized methodology for jamming detection based on a one-class classifier of black and white images extracted from raw I-Q samples using \emph{sparse autoencoders}.
    \item We experimentally compare our new methodology with those proposed in~\cite{alhazbi2023_ccnc} and~\cite{alhazbi2023_arxiv}, showing remarkable performances and improvements concerning the \ac{RJP} at the receiver, distance from the jammer, number of I-Q samples per image, training set size, and invariance to the hardware used for training.
    \item We provide additional results on a new dataset gathered using new hardware, namely, the LimeSDR.
    \item We provide new results on the data collected using various sampling rates at the jammer and receiver.
    \item We provide new results on using a new jamming strategy, i.e., deceptive jamming using the same modulation (\ac{BPSK}) and signal of the legitimate communication link. 
\end{itemize}

We acknowledge that \sol\ takes inspiration from anomaly detection strategies applied in other research domains, e.g., intrusion detection and computer vision~\cite{gong2019_iccv}. However, to the best of our knowledge, none of the contributions in the current literature provided a structured methodology to apply such strategies to detect jamming attacks on the wireless RF spectrum. Also, as described in more detail in Sect.~\ref{sec:related}, none of the contributions focused on low-BER regimes, investigating how to detect jamming while still maintaining remote communication capabilities. 

{\bf Roadmap.} The rest of this paper is organized as follows. Section~\ref{sec:preliminaries} introduces preliminary notions; Section~\ref{sec:sys_adv_model} describes the scenario and adversarial model; Section~\ref{sec:methodology} provides the rationale and details of \sol; Section~\ref{sec:results} discusses our extensive measurement campaign and performance assessment of our solution, and finally Section~\ref{sec:conclusion} draws the conclusion and outlines future work.

\textcolor{black}{
\section{Related Work}
\label{sec:related}
\begin{table*}[!h]
\color{black}
\centering
    \caption{Qualitative comparison of \sol\ with related literature on jamming detection. The symbol $\newmoon$ denotes that a specific feature is supported, the symbol $\LEFTcircle$ denotes that the specific feature is partially supported, while the symbol $\fullmoon$ denotes that the feature is not supported.
    }
    \label{tab:related}
\begin{tabular}{P{1.5cm}|P{3.5cm}|P{2.5cm}|P{2.2cm}|P{2.55cm}|P{2.5cm}}
\textbf{Ref.} & \textbf{\begin{tabular}[c]{@{}c@{}}Jamming Detection\\ Metric\end{tabular}} & \textbf{\begin{tabular}[c]{@{}c@{}}Jamming Detection \\ Technique\end{tabular}} & \textbf{\begin{tabular}[c]{@{}c@{}}Robustness to\\ Jamming Distance\end{tabular}} & \textbf{\begin{tabular}[c]{@{}c@{}}Robustness to \\ Jamming Signal Type\end{tabular}} & \textbf{\begin{tabular}[c]{@{}c@{}}Jamming Detection\\ in Low-BER Regime\end{tabular}} \\ \hline
\cite{saxena2022_pmc} & RSS & Geometric and Arithmetic Mean ratio & $\fullmoon$ & $\fullmoon$ & $\fullmoon$ \\
\cite{ccakirouglu2010_icst} & PDR & Query-based procedure & $\fullmoon$ & $\fullmoon$ & $\fullmoon$ \\
\cite{borio2015_icl} & Carrier-to-Noise density power ratio & Sum-of-Squares Paradigm & $\fullmoon$ & $\fullmoon$ & $\fullmoon$ \\
\cite{akhla2017_wlet}    & Coherence blocks & Generalized Likelihood Ratio Test & $\fullmoon$ & $\fullmoon$ & $\fullmoon$ \\
\cite{strasser2010_tosn} & RSS & Predetermined knowledge, Error Correcting Codes, and Limited node Wiring & $\fullmoon$ & $\fullmoon$ & $\fullmoon$ \\
\cite{chiang2007_mobicom}, \cite{chiang2011_ton} & PDR & Code Tree & $\fullmoon$ & $\fullmoon$ & $\fullmoon$ \\
\cite{marttinen2014_milcom} & Re-transmissions & Statistics-based & $\fullmoon$ & $\fullmoon$ & $\fullmoon$ \\
\cite{axell2015_navigation} & \ac{AGC} & Static tests & $\fullmoon$ & $\newmoon$ & $\fullmoon$ \\
\cite{punal2014_wowmom} & PDR & Random forests & $\fullmoon$ & $\fullmoon$ & $\fullmoon$ \\
\cite{liu2012_mass} & PDR & Channel probing & $\fullmoon$ & $\fullmoon$ & $\fullmoon$ \\
\cite{lu2014_tmc} & Retransmisions & Message Invalidation Ratio & $\fullmoon$ & $\newmoon$ & $\fullmoon$ \\
\cite{pawlak2021_wiseml} & OFDM parameters & \ac{ML} & $\fullmoon$ & $\fullmoon$ & $\fullmoon$ \\
\cite{swinney2021gnss} &  Power spectral density, spectrogram, raw
constellation & \ac{ML} & $\newmoon$ & $\fullmoon$ & $\fullmoon$ \\
\cite{li2022_access} & Spectrogram & \ac{ML} & $\newmoon$ & $\newmoon$ & $\fullmoon$ \\
\cite{lu2022_tim} & Nonlinear alternating current & CUSUM & $\newmoon$ & $\fullmoon$ & $\fullmoon$ \\
\cite{wang2022_hpsr} & PHY, Radio Link Control and Packet Data Convergence Control parameters & LSTM & $\fullmoon$ & $\newmoon$ & $\fullmoon$ \\
\cite{bouzabia2022_jsyst} & I-Q & Autoencoders & $\newmoon$ & $\fullmoon$ & $\fullmoon$ \\
\cite{alhazbi2023_arxiv} & I-Q & CNN & $\newmoon$ & $\fullmoon$ & $\LEFTcircle$ \\
\cite{alhazbi2023_ccnc} & I-Q  & CNN & $\newmoon$ & $\fullmoon$ & $\LEFTcircle$ \\ \hline
{\bf \sol} & I-Q & Sparse Autoencoders & $\newmoon$ & $\newmoon$ & $\newmoon$ 
\end{tabular}
\end{table*}
Several scientific papers recently considered drones for indoor applications, focusing on aspects such as localization~\cite{famili2022_tvt}, navigation~\cite{raja2021_tvt}, \cite{jung2018_ral}, and visualization~\cite{gao2020_sensorsj}. However, none of them investigates jamming attacks and anti-jamming approaches for indoor scenarios, thus mainly referring to the literature on generic (outdoor) jamming detection.\\
In the scientific community, jamming detection is usually achieved by applying various types of analysis on one or more metrics extracted from the primary communication link.\\%
Regarding the metrics, several parameters have been analyzed, such as the \ac{RSS} of the signals~\cite{saxena2022_pmc}, the \ac{PDR} as in~\cite{ccakirouglu2010_icst} and~\cite{punal2014_wowmom}, the Carrier-to-Noise density power ratio~\cite{borio2015_icl}, retransmission attempts~\cite{marttinen2014_milcom}, and~\cite{liu2012_mass}, the packet re-transmission profile, as in~\cite{lu2014_tmc}, or modulation-specific metrics, such as for \ac{OFDM} in~\cite{pawlak2021_wiseml}. Such metrics have been used in several scenarios and communication technologies, e.g., Massive MIMO~\cite{akhla2017_wlet}, \ac{WSN}~\cite{strasser2010_tosn}, GPS~\cite{axell2015_navigation}, IEEE 802.11~\cite{punal2014_wowmom}, and spread spectrum-based communication technologies~\cite{chiang2007_mobicom},~\cite{chiang2011_ton}.\\
At the same time, due to the increasing popularity of \ac{AI}, \ac{ML} and \ac{DL} approaches have been recently used extensively for detecting ongoing jamming. Such tools include \acp{CNN} such as in~\cite{swinney2021gnss} and~\cite{li2022_access}, genetic algorithm-based Cumulative Sum (CUSUM) methods such as in~\cite{lu2022_tim}, Bayesian networks such as in~\cite{wang2022_hpsr} and, finally, autoencoders such as in~\cite{bouzabia2022_jsyst}. All such approaches utilize as the main source of information the \ac{PHY} layer of the communication stack due to its direct relationship with the wireless channel, where jamming occurs. However, although some of the contributions cited analyzed the performance of the proposed jamming detection technique with low \ac{SNR}, none of them considered the \ac{BER} of the communication link. As a result, the proposed approaches mostly confirm that the root cause of the drop in \ac{BER} is jamming. However, they cannot detect such attacks even when the jamming effect is so low as not to significantly affect the \ac{BER} of the communication link. As explained above, such consideration is particularly relevant in mobile scenarios for remotely controlled equipment, not to lose control of the mobile entity completely before detecting jamming. In this context, the only contribution to the literature achieving such a property is our previous proposal in~\cite{alhazbi2023_ccnc}. As shown in Section~\ref{sec:results}, the methodology shown in this paper significantly outperforms both the solution proposed in~\cite{alhazbi2023_ccnc} and the improvements of such a methodology, as the one proposed in~\cite{alhazbi2023_arxiv}. We summarize our comparison with the current literature in Tab.~\ref{tab:related}.
}







\section{Preliminaries}
\label{sec:preliminaries}

In this section, we introduce preliminary notions that are useful to the readers of this manuscript, i.e., digital modulation techniques (Section~\ref{sec:modulation}) and autoencoders (Section~\ref{sec:autoencoders}).

\subsection{Digital Modulation}
\label{sec:modulation}

\begin{figure}
    \centering
    \includegraphics[width=\columnwidth]{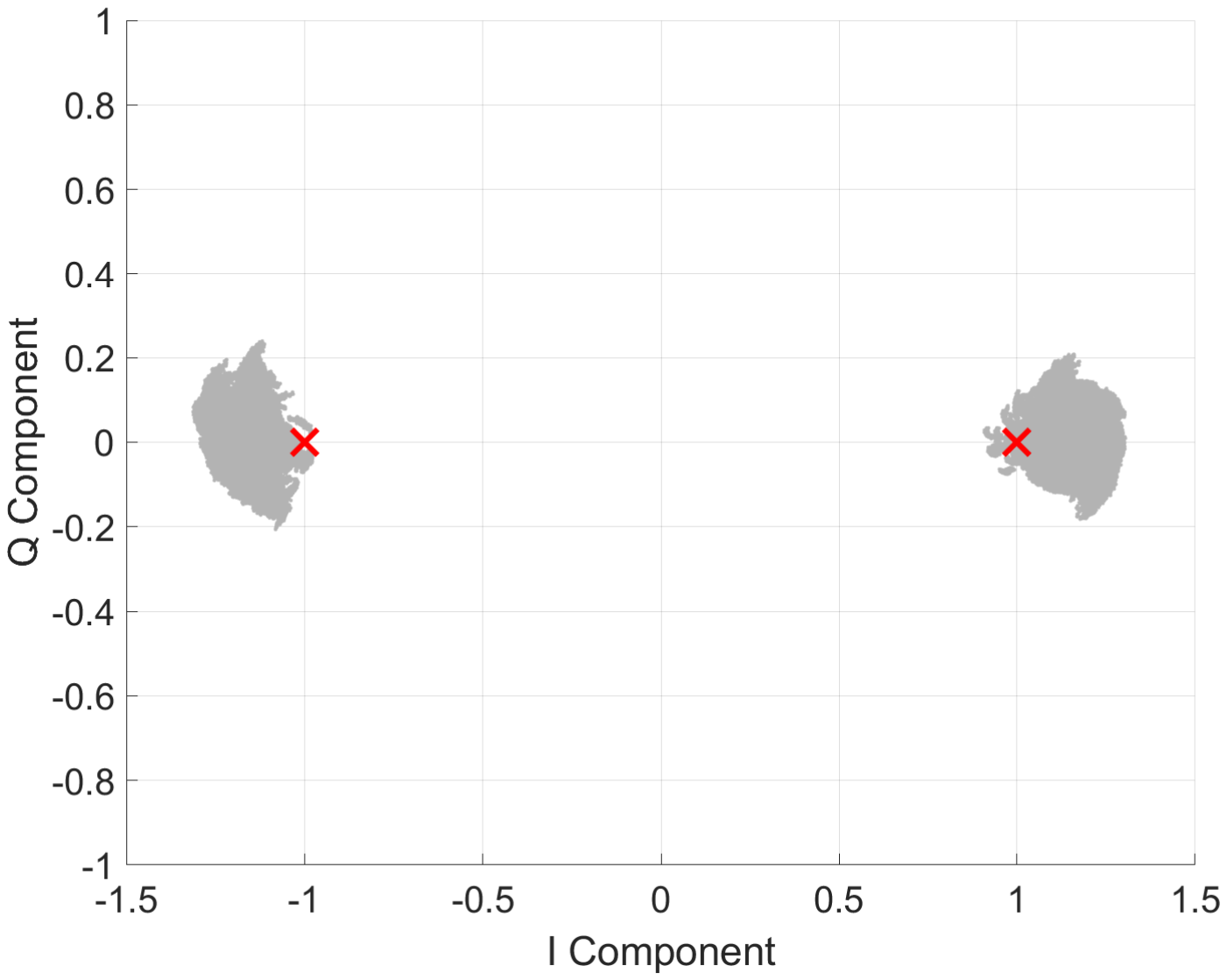}
    \caption{I-Q plane of a BPSK modulation. The receiver expects two symbols, i.e., $\left[ i=-1, q=0 \right]$ and $\left[ i=1, q=0 \right]$ (red crosses). However, due to the wireless channel, the received symbols are displaced (light grey area).} 
    \label{fig:iq_image}
\end{figure}

Digital modulation schemes adopted in wireless communication systems preprocess baseband signals to make them suitable for transmission at high frequencies~\cite{rappaport2001_book}. Typically, modulation techniques divide the bit-stream to be transmitted into two orthogonal components, namely the I vector and the Q vector, linked in a complex value of type $I+jQ$, where the I vector is the real component and the Q vector is the imaginary component. Due to their orthogonality, such components can be transmitted together on the wireless channel without interfering. They can also be recovered and assembled at the receiver to reconstruct the original bit-stream. 
In this context, a typical way to represent complex I-Q signals is through the I-Q plane, as shown in Fig.~\ref{fig:iq_image}. 
In particular, the number of expected I-Q values at the receiver (denoted as $n$) indicates the number of bits that can be recovered through a single complex I-Q value. In general, we can recover $\log_{2}{n}$ bits from $n$ symbols, and thus, with reference to Fig.~\ref{fig:iq_image} showing the I-Q plane of a \ac{BPSK} modulation, we can recover $n=2$ bits, i.e., $\left[ i=-1, q=0 \right]$ (b=0) and $\left[ i=1, q=0 \right]$ (b=1).
We also notice that the values of the I-Q samples at the receiver always differ from those at the transmitter because of noise introduced by the hardware components of the devices and the fluctuations of the wireless channel between the transmitter and the receiver. To recover the original transmitted symbol, the receiver associates the expected symbol with the received I-Q sample whose distance from the received one is the shortest. Thus, the higher the noise impact, the higher the chance that a received I-Q sample is associated with the wrong expected symbol, leading to an error and, therefore, a higher \ac{BER}. At the same time, with a given noise profile affecting the communication channel, the higher the modulation order $n$, the higher the amount of expected symbols and, thus, the higher the \ac{BER}. The rationale described above is adopted by lower-order modulation schemes, e.g., communication links affected by high noise levels, such as satellite transmissions and mobile indoor applications.
The intuition driving our work is that the collective displacement of I-Q samples from the expected one can be used to discriminate the presence of various levels of intentional interference, i.e., jamming, affecting the communication link. We will provide more details about our approach in Section~\ref{sec:methodology}.

\subsection{Autoencoders}
\label{sec:autoencoders}

Without loss of generality, \emph{autoencoders} are a special type of \ac{ANN} which can be trained to reconstruct their input~\cite{bank2021_arxiv}. Formally, the problem autoencoders solve is to find an encoder $A: \mathbb R^{d} \to \mathbb R^p$ and decoder $B: \mathbb R^p \to \mathbb R^d$ satisfying Eq.~\ref{eq:autoenc_eq}.
\begin{equation} 
    \label{eq:autoenc_eq}
    \arg \min _{A, B} E[\Delta (\mathbf x, B \circ A (\mathbf x))],
\end{equation}
where the symbol ``$\circ$" represents the composition operator, i.e., $B \circ A( \mathbf x) = B(A(\mathbf x))$, $E$ represents the expectation of the distribution of the input $\mathbf x$, '$A (\mathbf x)$' the encoded version of the input, known as the \emph{bottleneck} of the autoencoder when $p < d$, and finally $\Delta$ is the reconstruction loss function, which measures the distance between the input of the \ac{ANN} and the reconstruction of the input \cite{bank2021_arxiv}. For our purposes, in line with many scientific contributions such as~\cite{bank2021_arxiv} and~\cite{oligeri2022_tifs}, $\Delta$ is the \ac{mse} function, as defined in Eq.~\ref{eq:mse_eq} on two reference distributions $\mathbf x$ and $\mathbf y$.
\begin{equation} 
    \label{eq:mse_eq}
    \text{mse}(\mathbf x, \mathbf y) = \frac{1}{d} \cdot \| \mathbf x - \mathbf y \|^2_2 \text{ for all } \mathbf{x}, \mathbf{y} \in \mathbb{R}^{d},
\end{equation}
being $d = M \cdot N$.

Traditionally, autoencoders have been used mainly for image generation, particularly for creating sets of images similar to the input ones. However, in the cybersecurity research domain, they are mainly used for anomaly detection. Let $c$ be an autoencoder trained on samples from a probability distribution $P$. Next, let $Q$ be a probability distribution such that $P \ne Q$. Then, we expect $c$ to have a smaller reconstruction error when tested on unseen samples from $P$ than when tested on unseen samples from $Q$. Therefore, the magnitude of the reconstruction error of $c$ in an unseen sample measures the probability that such an unseen sample is not sampled from $P$ but from another distribution~\cite{oligeri2022_tifs}. Consequently, we can define a specific error value $\tau$ as a classification boundary. All samples in which the auto-encoder $c$ makes a higher error than $\tau$ can be classified as 'not from $P$'. In the literature, $\tau$ is often referred to as a \emph{threshold}~\cite{erfani2016_patrec, khan2017_esa}.

In this work, we use autoencoders to build a statistical profile of the channel experienced between the mobile transmitter and the receiver under regular operating conditions. We provide more details in Section~\ref{sec:methodology}. 
 
\section{System and Adversary Model}
\label{sec:sys_adv_model}

\begin{figure}
    \centering
    \includegraphics[width=\columnwidth]{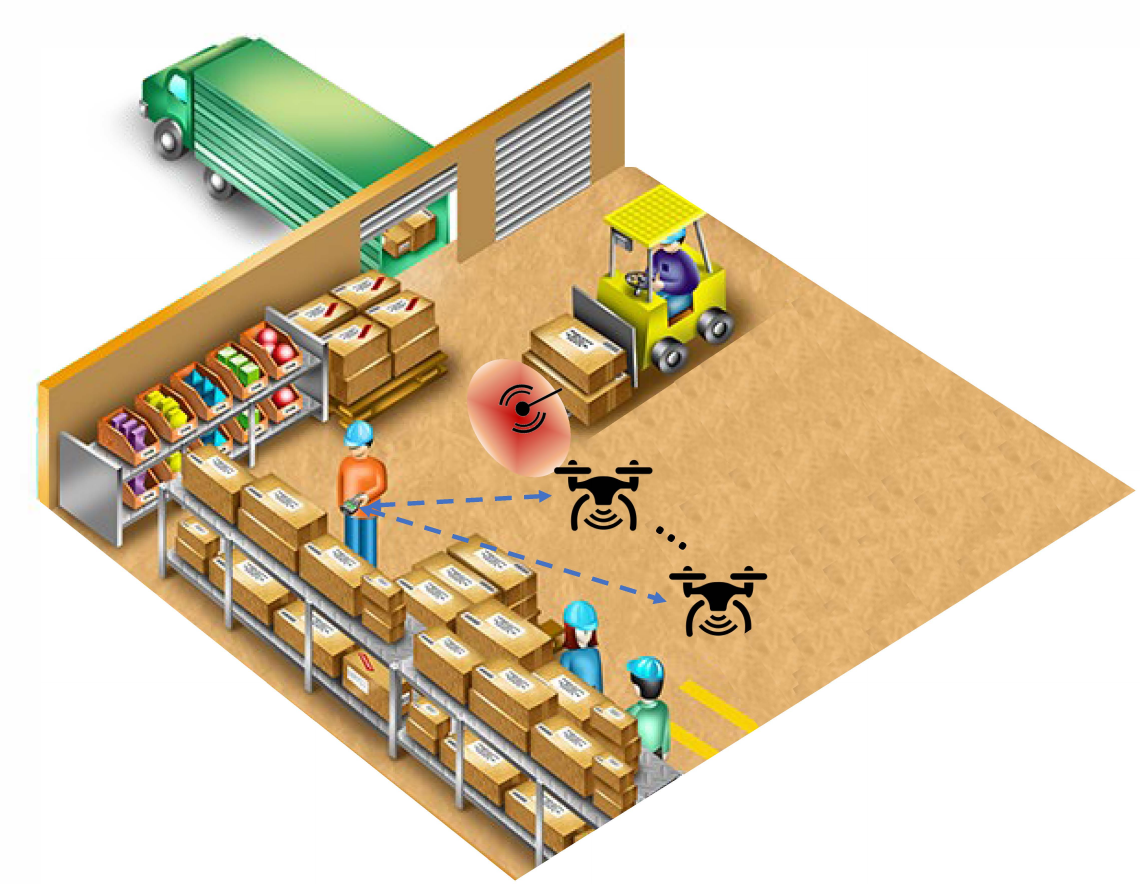}
    \caption{\textcolor{black}{Reference Scenario. While operating indoors, a drone tries to detect jamming (red color) in a low-BER regime, i.e., before the jamming affects the quality of the legitimate communication link (blue color). } 
    }
    \label{fig:scenario} 
\end{figure}

Fig.~\ref{fig:scenario} shows the scenario and adversary model considered in this work, inspired by the real-world use case discussed in~\cite{ikea_drones}.
\textcolor{black}{We consider an indoor scenario where drones operate to achieve (semi)-automatized tasks, e.g., warehouse inspection or home surveillance. The relevance of such a scenario is confirmed by recent news, reporting that leading companies such as IKEA and Amazon are already experimenting with commercial products for this purpose~\cite{ikea_drones},~\cite{amazon_indoorDrone}.} We consider the existence of a communication link between the drone and a \ac{GCS}, which can be used either to pilot the drone, in the case of a \ac{RPAS}, or to report telemetry data, in the case of (semi) autonomous operations~\cite{tedeschi2023_acsac}. Independently of the communication link usage, we do not make any assumption on the nature of the drone operations, which can be either instructed by a human pilot or (semi-) autonomous. Also, we do not make assumptions on the presence of a positioning technology, besides that it is not affected by spoofing~\cite{raponi2021_sp}. Without loss of generality, we consider that the communication link between the drone and the \ac{GCS} adopts the \ac{BPSK} modulation scheme. The mentioned assumption is reasonable, as such a scheme allows one to mitigate the noise affecting indoor communication channels as much as possible, being also used in modern WiFi standards.

We also consider the deployment of a static jammer in the area, which injects noise into the wireless channel used for communication between the drone and the \ac{GCS}. We assume that such a jammer continuously emits interfering signals with the highest possible transmission power, to affect ongoing wireless communications in the deployment area as much as possible. Unlike the contribution in~\cite{alhazbi2023_ccnc}, we do not make any assumptions about the specific jamming signal: it can be \ac{AWGN}, a single tone, and even deceptive jamming, adopting the same modulation scheme used by the legitimate communication link. Also, being possibly unaware of the sampling rate of the legitimate communication link, the jammer transmits signals with the highest possible sampling ratio, limited only by the hardware used to carry out the attack. Note that when the attacker is unaware of the modulation used by the legitimate communication link, they can perform modulation-agnostic jamming, e.g. by injecting \ac{AWGN} or a single tone centered on the channel of interest. Instead, suppose that the attacker is aware of the modulation used by the target link. In that case, it can use \emph{deceptive jamming}, i.e., injecting a signal characterized by the same digital modulation (and possibly the same message pattern) of legitimate messages, further complicating the detection process.

Being bounded by the maximum achievable transmission power, the jammer significantly impacts ongoing communications only in a specific area around its location. In fact, the \ac{RSS} associated with the jamming signal depends on the distance between the jammer and the receiver location (drone), which is highest in the proximity of the jammer and decreases with further movement. The described wireless propagation effect generates a \emph{jammed area}, disrupting wireless communications. In this context, the mobile receiver (drone), moving toward the jammed area, wants to promptly detect the jamming signal in a low-BER regime, i.e., before the effect of the jamming on the quality of the communication link becomes noticeable, causing a significant increase in the \ac{BER}. In fact, such a jamming detection mechanism would improve drone situational awareness, as it would allow \ac{GCS} to be aware of imminent jamming and take action immediately without relying on a predefined set of steps (e.g., landing, returning to the starting point). 

\section{Methodology}
\label{sec:methodology}

This section describes \sol, i.e., the methodology we propose to detect jamming in a low-BER regime. In summary, \sol transforms jamming detection into an anomaly detection problem on images generated by encoding the current state of the communication channel. Overall, we can identify two main building blocks of our solution: the \emph{Image Generation} and the \emph{Jamming Detection}, described below. Tab.~\ref{tab:notation} summarizes the primary notation used below, with a short description.
\begin{table}[!t]
    \centering
    \caption{Notation and brief description.
    }
    \label{tab:notation}
    \begin{tabular}{P{1.2cm}|P{6.2cm}}
         {\bf Notation} & {\bf Description}  \\ \hline
         n & Number of I-Q samples per image. \\
         M, N & Dimensions of the input image. \\
         $a_{m,n}$ & Generic pixel of the input image. \\
         $d$ & Dimension of vectors within autoencoders, with $d=M\cdot N$. \\
         K & Encoder units. \\
         J & Decoder units. \\
         $\tau$ & Autoencoder threshold value. \\
         $MSE_{train}$ & \ac{mse} value obtained at training time. \\
    \end{tabular}    
\end{table}

\textcolor{black}{
\textbf{Image Generation.} Fig.~\ref{fig:img_gen} provides a graphical overview of the image generation process used in \sol.
\begin{figure*}
    \centering
    \includegraphics[width=\textwidth]{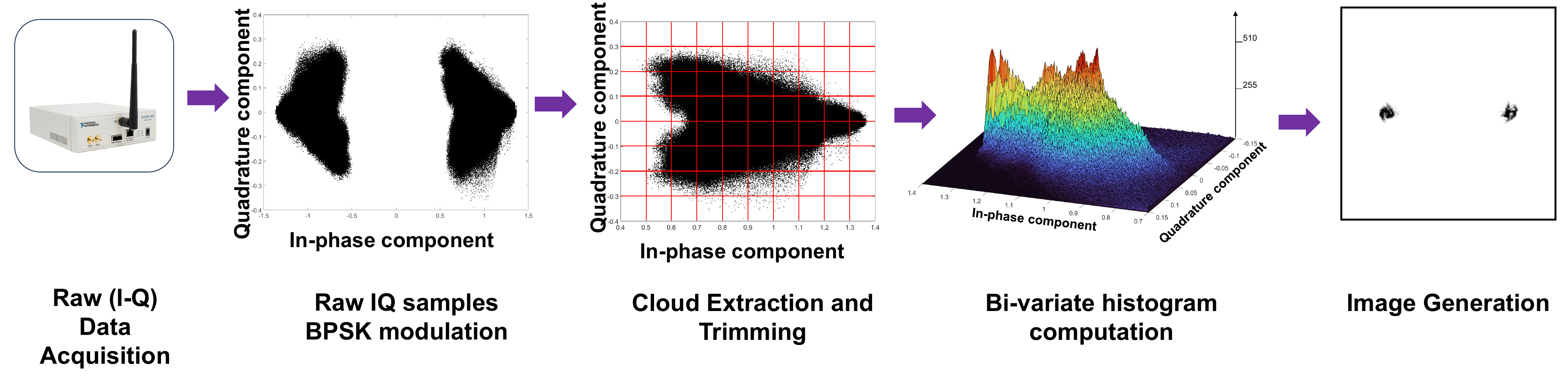}
    \caption{\textcolor{black}{Graphical overview of the image generation process of \sol. We acquire raw I-Q data through a generic \ac{SDR}, plot them through the traditional I-Q plane representation, compute a bi-variate histogram based on the density of samples in given areas of the plane, and then obtain an image.}} 
    \label{fig:img_gen}
\end{figure*}
The input to the process is represented by a sequence of raw I-Q samples. Such samples can be collected using a \ac{SDR} or any hardware capable of obtaining \ac{PHY} wireless channel information (e.g., spectrum analyzers). The amount of samples used to generate images, namely $n$, is one of the degrees of freedom of our solution and can be configured by trading off the overall accuracy with the computational requirements of the solution (see Section~\ref{sec:results} for a detailed evaluation of the impact of this parameter).
We represent the sequence of I-Q samples through the traditional I-Q plane, where we display the component $I$ on the x-axis and the component $Q$ on the y-axis. In a benign scenario (no jamming), such a representation generates several clouds of I-Q values, approximately centred on the value of the expected \emph{symbol}, as explained in Section~\ref{sec:modulation}.
Based on this representation, we build a bi-variate histogram. Specifically, we divide the I-Q plane around the cloud of points into tiles $N \times M$, where the values of $N$ and $M$ depend on the dimensions of the images we want to obtain. 
Then, for each tile $a_{m,n}$, we evaluate the number of I-Q samples that fall into the tile itself. When the number of I-Q samples falling on a tile exceeds the value $255$, we truncate it to the maximum value, to ensure $a_{m,n} \in [0,255], \forall (m,n)$.
We consider the output of such a process as a pixel value. As a consequence, the output of the image generation process is a grayscale image corresponding to the received profile of I-Q samples.
Note that, in principle, we might also work with colored images (3-D matrices). In Section~\ref{sec:results}, we evaluate this configuration, adopted in~\cite{alhazbi2023_arxiv}, and show its pros and cons in physical-layer jamming detection. 
}

\begin{figure*}
    \centering
    \includegraphics[width=\textwidth]{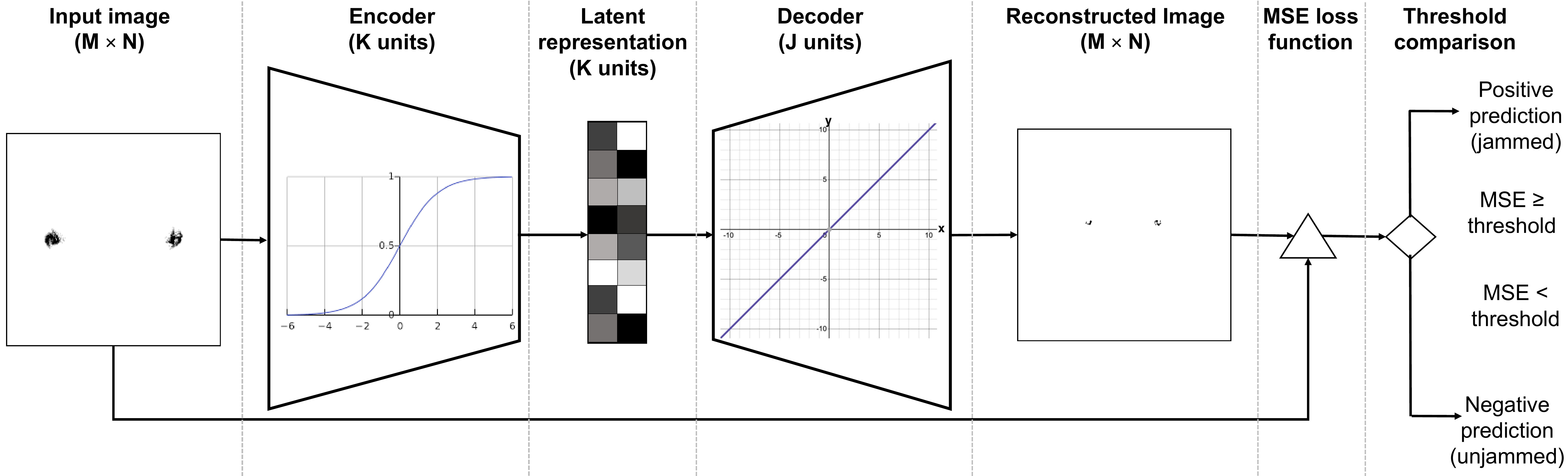}
    \caption{\textcolor{black}{Autoencoder architecture. We use a \emph{logsig} encoder transfer function and a \emph{purelin} decoder transfer function, with a total number of two hidden layers, a sparsity regularization term of 0.05, and the \ac{mse} as a loss function, coming up with a \emph{sparse nonlinear autoencoder}. 
    }
    }
    \label{fig:autoencoder}
\end{figure*}

\textbf{Jamming Detection.} \textcolor{black}{The jamming detection process is the building block of \sol\ dedicated to the timely detection of possible jamming affecting the wireless communication channel. It is a \ac{DL}-based process using \emph{sparse autoencoders}, so it involves a \emph{training} and \emph{testing} process. We highlight that our manuscript does not aim to provide a new autoencoder architecture. Instead, as discussed in the related work (Sect.~\ref{sec:related}) and summarized in Tab.~\ref{tab:related}, the innovation of our manuscript is that \sol\ is the first solution applying autoencoders to solve a jamming detection problem.}
Fig.~\ref{fig:autoencoder} shows the architecture of the adopted \emph{autoencoder}.

\textcolor{black}{The input to the autoencoder is represented by the images obtained as a result of applying the \emph{image generation} process to I-Q samples collected from the wireless channel. At training time, we acquire I-Q samples corresponding to the typical behavior of the communication link. We denote images corresponding to such a scenario configuration as \emph{unjammed images}. We acquire $n$ I-Q samples and generate images of size $M \times N$. In our deployment, $n=10^{5}$ and $M = N = 224$, to match the size of images used as input to various \ac{NN} (we compare our performance to two benchmark solutions using \ac{CNN} in Section~\ref{sec:results}). 
As encoder, we used logarithmic sigmoid functions with $K=16$ units (a.k.a. neurons). As a result, we obtain a latent representation of the input image consisting of $K=16$ dimensions. Such a latent representation summarizes the relevant features of the input image, significantly compressing its dimension (compared to the input image).
Then, we submit the latent representation vectors to the decoder using a linear decoder transfer function with a total number of $J = 50,176$ neurons. In our context, using the logistic sigmoid activation function in the decoder's units did not allow our solution to converge to a good solution. We suspect \textit{vanishing gradients} to be the cause, further supported by the better convergence we obtain when using linear activation functions in our decoder layer. 
In principle, autoencoders allow the use of multiple hidden layers. Overall, the higher the number of hidden layers, the better the performance of the classifier at run-time, but also the higher the computational overhead of the methodology. Here, we use two hidden layers and the sparsity regularization technique, in line with the architecture of \emph{sparse nonlinear autoencoders}. As we show in Section~\ref{sec:results}, such a choice allows us to obtain remarkable classification accuracy while achieving a computational cost lower than that of more complex architectures.}
\textcolor{black}{This process provides a reconstructed image of the same dimension as the input. We first convert the matrix of size $m \times n$ into a vector of dimension $d = M \cdot N$, concatenating the rows of the image one after the other. Then, we compute the \ac{mse} loss function as in Eq.~\ref{eq:mse_eq} (see Section~\ref{sec:autoencoders}). During training, we acquire several images corresponding to the regular (expected) behavior of the wireless channel, building a corresponding profile of such a channel when displayed through images (i.e., our hypothesis). We compute a threshold $\tau$ on such a profile, as explained in Section~\ref{sec:autoencoders}, to distinguish the \emph{regular channel conditions} from the unexpected one. At testing time, we compare the \ac{mse} value obtained from a run-time acquisition of the wireless channel with the threshold $\tau$ previously cited. Suppose that the \ac{mse} of the input image is equal to or greater than the threshold. In that case, the autoencoder produces a \emph{positive prediction}, meaning that a jammer affects the communication channel. Otherwise, when \ac{mse} is lower than the threshold, the autoencoder outputs a \emph{negative prediction}, meaning there is no jamming. In line with the logic of any \ac{ML} and \ac{DL} solution, the rationale of \sol\ aligns with statistical hypothesis testing methods.} In the following, we provide more details on the autoencoder training process and the threshold selection methodology.

{\bf Training the autoencoder.} For training the autoencoder, we use only \emph{unjammed images}, i.e., images generated from I-Q samples acquired when no jamming affects the communication link. We do not use any \emph{jammed images}, i.e., images obtained from a jammed communication channel. We applied this strategy mainly since jamming can be performed in many different ways, typically unknown and unpredictable to the legitimate parties. Instead, our intuition is that we can build a profile of the expected conditions of the communication link, even in very noisy scenarios, and detect jamming as a deviation from such expected conditions.
In this context, to guarantee reliable operations for the autoencoder, it is crucial to gather I-Q samples that cover the most extensive possible set of expected conditions of the communication channel. In fact, the reliability of the auto-encoder in identifying anomalies leading to jamming depends on the variety of conditions affecting the communication channel, thus reducing false positive events.

\begin{figure}
    \centering
    \includegraphics[width=\columnwidth]{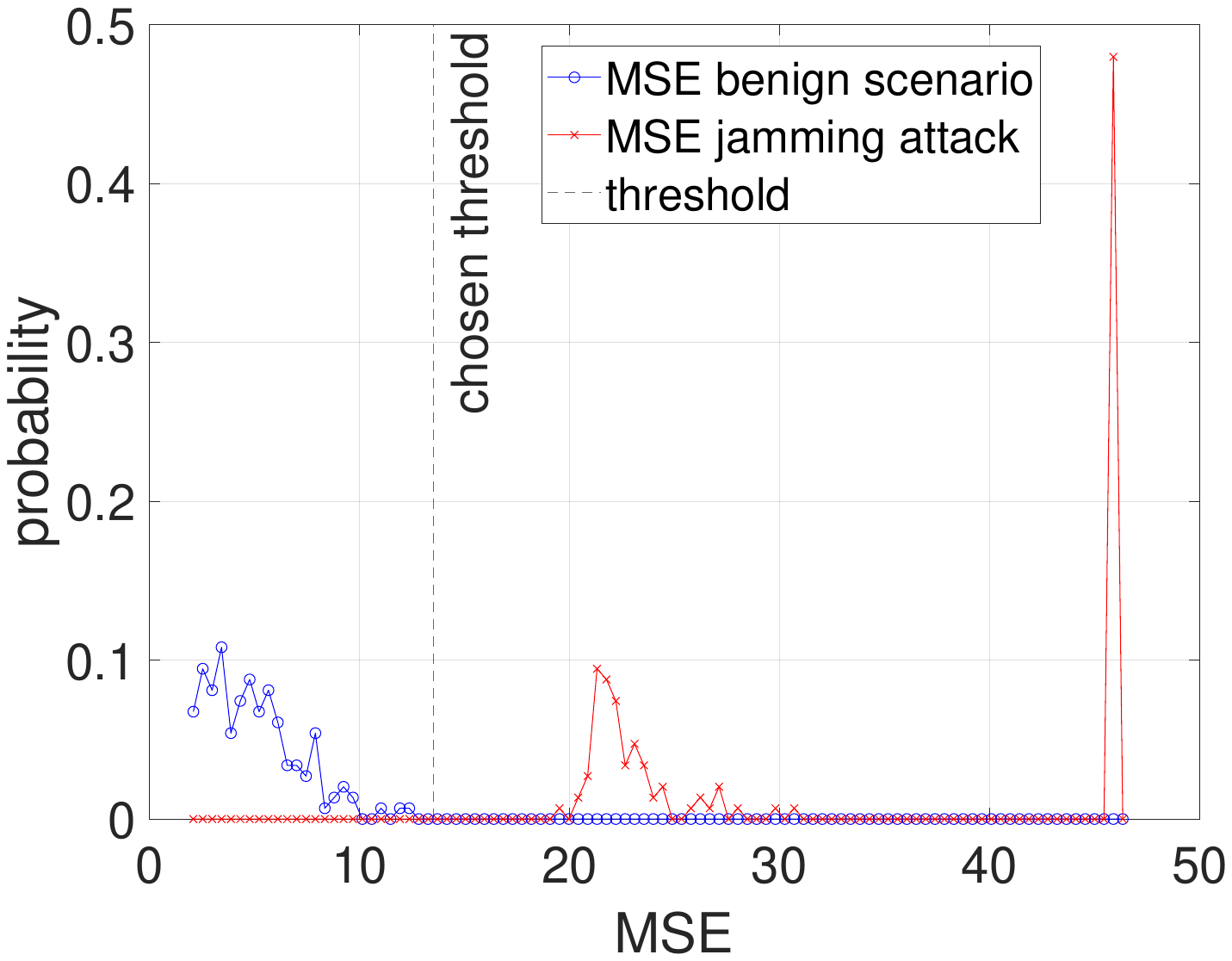}
    \caption{Sample distribution of the \ac{mse} of unjammed and jammed images and optimal threshold selection. The reported data refer to a jammer jamming with $RJP=0.6$ at a distance of $10$~meters from the receiver.} 
    \label{fig:mse_example}
\end{figure}

{\bf Threshold selection.} Optimal selection of the decision threshold of an autoencoder is an actively researched problem, which has yet to have a universally optimal solution \cite{torabi2023_cybersecurity, khan2017_esa}. In this work, we adopt the approach suggested by the authors in~\cite{erfani2016_patrec}, i.e., we compute the threshold according to Eq.~\ref{eq:threshold_eq}.
\begin{equation} 
    \label{eq:threshold_eq}
    \tau = {mean} ({MSE}_{\text{train}}) + 3.5 \cdot {std} ({MSE}_{{train}}),
\end{equation}
being ${MSE}_{{train}}$ the set of \acp{mse} that the autoencoder compute on the training data, ${mean}$ the statistical average and ${std}$ the standard deviation.
As demonstrated by the authors in~\cite{erfani2016_patrec} and confirmed by the authors in \cite{khan2017_esa}, such a choice is reasonable in scenarios where no anomalous samples are used in the training phase, as in our scenario. Furthermore, as acknowledged by the authors in~\cite{khan2017_esa}, this choice reduces false negatives compared to the standard option, i.e., setting $\tau$ to the maximum \ac{mse} observed in the training samples. In turn, such a choice increases the chances of detecting jamming in a low-BER regime. Figure~\ref{fig:mse_example} reports an example of the threshold selection process on actual data acquired with jammer jamming with $RJP=0.6$ at a distance of $10$~meters from the receiver (see Section~\ref{sec:measurements} for details). We notice that the distribution of the \ac{mse} values for unjammed images is characterized by smaller values compared to the one of jammed images, with only minimal overlap at the tails of the distributions. Setting the threshold according to Eq.~\ref{eq:threshold_eq} allows one to reduce false negatives without affecting performance.

Finally, note that deploying an autoencoder for image-based jamming detection involves setting optimal values of the hyperparameters of such a tool. We discuss the selection of auto-encoder hyperparameters in Section~\ref{sec:results}.

\section{Experimental Assessment}
\label{sec:results}

In this section, we provide the details of our extensive experimental assessment, carried out to evaluate \sol\ in real indoor scenarios. We introduce the actual measurements used for the following analysis in Section~\ref{sec:measurements}, while in Section~\ref{sec:exp_settings} we describe the experimental settings. Then Section~\ref{sec:autoenc_robustness} reports the performance of our approach and compares it with two solutions available in the literature. We extend such results further, evaluating \sol\ with different hardware and various channel sampling rates (Section~\ref{sec:samplingRate}) and investigating the capability of identifying deceptive jamming (Section~\ref{sec:deceptive}).

\subsection{Measurements}
\label{sec:measurements}

In this paper, we build on top of the data provided as part of our contribution in~\cite{alhazbi2023_dib} and extend such a dataset with new measurements obtained with new hardware and different configuration parameters.

All the measurements discussed below have been acquired in an indoor office environment during a regular working day, with people moving around and possibly across the legitimate communication link. Such conditions match, as much as possible, those described in Section~\ref{sec:sys_adv_model}.

The general setup of our measurements includes three entities, i.e., a transmitter, a receiver, and a jammer. In the first set of experiments, we considered \ac{SDR} Ettus Research X310~\cite{ettus} featuring a daughterboard UBX160 as reference hardware for all the entities involved in our measurement campaign. Here, we placed the transmitter and the jammer close to each other while moving the receiver at various distances from the transmitter (see below).
For the second set of experiments, we used different hardware, i.e., the LimeSDR~\cite{limesdr}. It is a low-cost, open-source, \ac{SDR} platform supporting any wireless communication standard. For these experiments, we placed the transmitter and receiver $3$ meters away from each other and placed the jammer between them at a distance of $1.5$~meters from both entities.
We connected the \acp{SDR} either via Ethernet (Ettus X310) or USB 2.0 (LimeSDR) to two laptops, one controlling the data transmission and jamming processes and the other taking care of data reception. The received I-Q samples were stored on the laptop connected to the receiving \ac{SDR} and subsequently uploaded to a centralized server for data analysis. Specifically, we used the High-Performance Computing (HPC) cluster available at TU/e, Eindhoven, The Netherlands, providing a CPU E2124 with four cores running at 3.3 GHz and 32 GB of RAM, as well as 2 GPUs Tesla V1000 running at 32 with 256 GB of RAM. 

Regarding software, for both setups, we used the GNURadio v3.8 development toolkit~\cite{shea2016_grc}. We set the carrier frequency $f_c = 900$ MHz for both the legitimate communication link and the jamming. We configured the transmitter and receiver to exchange packets containing a repeating sequence of 256 bytes, encoded by a \emph{Constellation Modulation} block using the regular \ac{BPSK} modulation scheme.

For the first setup using the \ac{SDR} Ettus X310, we configured a sample rate of 1M samples per second at the transmitter, the receiver, and the jammer. We set the normalized transmission power and receiver gain to the maximum value of $1$, corresponding to approximately 15 dBm (32 mW) of transmission power. At the receiver, we set up the reception chain of the \ac{BPSK} modulation, including (i) an \emph{Adaptive Gain Control} (AGC) block, to mitigate the signal level fluctuations introduced by the multipath fading; (ii) a \emph{Symbol Sync}, which performs timing synchronization; (iii) a \emph{Costas Loop}, which locks to the center frequency of the signal and down-converts it to baseband; and, (iv) finally, a \emph{Constellation Decoder} block, which decodes the constellation points. We saved the I-Q data obtained as the output of the \emph{Constellation Decoder} block. We did not use any channel estimation techniques to filter out any channel effects beneficial for jamming detection. 
Regarding the jammer, we chained two blocks: (i) a signal source, which can be an analog $\sin$ wave (tone jammer) or a digital sequence of Gaussian-distributed values (Gaussian jammer); and (ii) the \emph{USRP Sink block}, which sends the signal to the radio for actual transmission. 
To emulate the scenario described in Section~\ref{sec:sys_adv_model}, we placed the entities at different distances and, to further mimic the movement, we changed the relative values of the jammer transmission power between $0$ and $0.8$, i.e., between 0 and 7.94mW (9dBm), respectively. 
Values greater than $0.8$ cause a complete disruption of the \ac{BER} of the signal (see Section~\ref{sec:autoenc_robustness}), making our solution unnecessary. In fact, when using a static setup, the reduction of the transmission power of the jammer makes the received jamming power level at the receiver weaker, allowing us to investigate the effect of a jammer located further away from the communication link.

For the second setup using the LimeSDR, we considered different sample rate values at the receiver and the jammer. In particular, consider the formula $t_s = K \times t_{S,R}$, where $t_{S,R}$ is the reference sample rate of 1M samples per second, $t_s$ is the actual sample rate used in the measurement, and $K$ an oversampling factor. We carried out experiments considering different values of the oversampling ratio both at the receiver and at the jammer, namely, the \emph{\ac{ROR}} and \emph{\ac{JOR}}. Specifically, we varied both the \ac{ROR} and the \ac{JOR} in the range $[1,4]$. For all such experiments, we tested two jamming strategies: tone jamming, i.e., jamming with a sinusoid signal, and deceptive jamming, i.e., jamming with precisely the same bit-sequence \ac{BPSK}-modulated signal delivered between the legitimate transmitter and the receiver. 

In general, the two measurement setups described above allowed us to investigate the effectiveness of \sol\ while varying an extensive range of configuration parameters, including: (i) the \ac{RJP} at the receiver, (ii) the distance of the jammer from the legitimate communication link, (iii) the oversampling ratios at the receiver and the jammer, and (iv) the types of jamming and the type of radios used for the experiments.

\subsection{Experimental Settings}
\label{sec:exp_settings}

For our experiments, we found the best configuration of the hyperparameters of the autoencoder of \sol\ and compared its performance with the approach based on the binary \ac{CNN} \emph{Resnet-18} used in~\cite{alhazbi2023_ccnc} and the solution based on 3-D images proposed in~\cite{alhazbi2023_arxiv}. 

{\bf Autoencoder deployment.} We fine-tuned the hyperparameters of the autoencoder used in \sol to find the best trade-off between classification accuracy and the general validity of the solution, i.e., to avoid overfitting. Specifically, we used the Matlab-provided implementation of autoencoders, version R2022b. As mentioned in Section~\ref{sec:methodology}, we used the \emph{Pure Linear} (\emph{purelin}) transfer function as the decoder transfer function. 
We used a formal hyperparameter optimization method performed on a subset of our dataset for all remaining hyperparameters of the autoencoder. Specifically, we selected $1,500$ images from I-Q samples acquired in the scenario with the most data available, i.e.,  with the receiver positioned $10$ meters from the transmitter and the jammer emitting jamming with a \ac{RJP} of $0.5$. We evaluated all combinations of the following hyperparameters: (i) hidden size (i.e., the size of the latent representation), with considered values being $8$, $16$, $32$ and $64$; (ii) sparsity regularization, with considered values being $1$, $0.5$, and $0$; (iii) L2-regularization term, with considered values being $0.01$, $0.001$, and $0.0001$; and finally, (iv) encoder transfer function, with considered values being \emph{logsig} and \emph{satlin}. Finally, we set all remaining hyperparameters to their default values provided by Matlab, specified at~\cite{trainAuto}. Such various configuration parameters led us to test a total number of $72$ configurations. For such tests, we used the methodology described in the following, inspired by the $k$-fold cross-validation technique used by the authors in~\cite{shvetsova2021}.
\begin{itemize}
    \item We divide all jammed images in the selected dataset into $10$ disjoint subsets of equal size.
    \item We divide all unjammed images in the selected dataset into evenly sized disjoint subsets of $10$.
    \item For each combination of hyperparameters considered $\alpha$, for every $i \in \{ 1, \dots, 10 \}$, we do the following:
    \begin{itemize}
        \item We train an autoencoder on all subsets of unjammed images, except for the $i$-th one. During training, we use the hyperparameters in $\alpha$.
        \item We compute the \ac{mse} values of the autoencoder on the images in the $i$-th subset of unjammed images.
        \item We compute the \ac{mse} values of the autoencoder on the images in the $i$-th subset of jammed images.
        \item Using the \ac{mse} values collected as part of the two steps above, we compute the \ac{AUC} of the \ac{ROC} curve corresponding to the autoencoder.
    \end{itemize}
    \item For each combination of hyperparameters $\alpha$, we take the average of the \ac{AUC}-values found in the iterations of the last cited step.
    \item Finally, we pick the combination of hyperparameters for which the last step mentioned above yielded the largest result.
\end{itemize}
Note that, like the authors in \cite{shvetsova2021}, we selected the \ac{AUC} as the optimization metric since it measures the quality of a classifier independently of the threshold selection process. Following such a hyperparameter selection process, we set the hidden size value to $16$, the sparsity regularization term to $0.5$, the L2-regularization term to $0.01$ and the encoder transfer function to the Logistic Sigmoid (\emph{logsig}). 
We also selected the number of epochs by examining how many iterations the autoencoder takes to reliably converge in the worst case, i.e., when setting the size of the latent representation to 64. We empirically established that this would occur after 250 epochs, and thus, we selected such a value to trade-off between classification accuracy and training time.
We use such a hyperparameter configuration to train the autoencoder of \sol\ for all the results discussed below on unseen data to avoid overfitting.

{\bf Benchmark Approaches.} For all experiments, we compare \sol\ with two benchmark solutions, i.e., the former version of \sol\ published in~\cite{alhazbi2023_ccnc} and the approach proposed by the authors in~\cite{alhazbi2023_arxiv}. We selected such solutions as they work successfully on \ac{PHY} data, i.e., I-Q samples, significantly outperforming other solutions in challenging scenarios, such as the one considered in this manuscript. 
Both approaches use the residual \ac{CNN} \emph{Resnet-18}, pre-trained on the \emph{ImageNet} dataset~\cite{ILSVRC15} and with the necessary modifications to the output layer necessary to fit the nature of the jamming detection problem. Specifically, we modified the output layer of \ac{CNN} to consider two possible classes as output, i.e., either \emph{No-Jamming} or \emph{Jamming}. Regarding the input layer, CNN works on images of size $224 \times 224$ constructed over I-Q samples collected from the wireless channel, which is optimal for comparison to \sol. To train CNN, we used the automated procedure \emph{trainNetwork} provided by Matlab. We set the batch size to $32$, the number of epochs to $1$, and the solver to \emph{adam}, similar to \cite{alhazbi2023_ccnc}. We set all remaining hyperparameters to their default values, available at~\cite{trainCNN}. 

\begin{figure}
    \centering
    \includegraphics[width=\columnwidth]{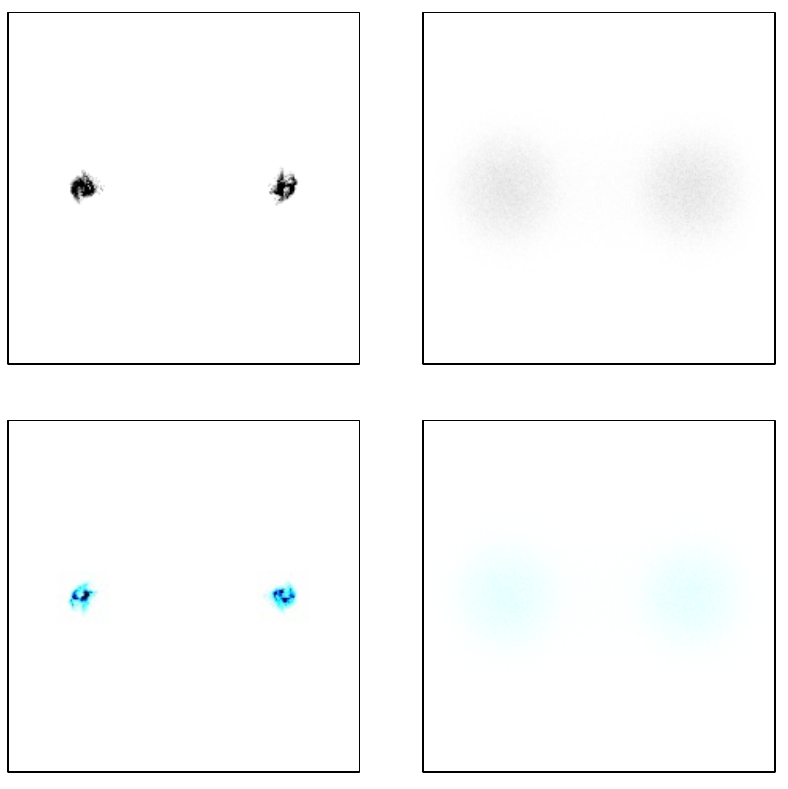}
    \caption{Sample grayscale and colour \emph{unjammed} (left) and \emph{jammed} (right) images generated from the same set of I-Q samples, used for the experimental assessment. We generated the \emph{jammed} images using samples collected at $10$ meters from the jammer, with \ac{RJP} = $0.5$.} 
    \label{fig:images}
\end{figure}

\textcolor{black}{The solution provided in~\cite{alhazbi2023_ccnc} works on grayscale images, while the proposal in~\cite{alhazbi2023_arxiv} uses color images characterized by three layers rather than one. We follow the same procedure described in~\cite{alhazbi2023_arxiv} to set up such images. As an example, Fig.~\ref{fig:images} compares a grayscale and a color image generated over the same set of I-Q samples collected from the wireless channel. 
The lower images in Fig.~\ref{fig:images} have been generated by considering one layer for each primary
colour component (red, green, and blue). Therefore, assuming an
image constituted by a three-layer matrix, i.e., [224 × 224 × 3] (one
layer for each primary colour), and the pixel value between 0 and
255, in line with~\cite{alhazbi2023_arxiv}, we assign each value of the tile through the following rule.
\begin{itemize}
    \item $0 \le x_T \le 255$, then $p_R = 0, p_G = 0, p_B = x_T$,
    \item $256 \le x_T \le 511$, then $p_R = 0, p_G = x_T - 255, p_B = 255$,
    \item $x_T > 511$, then , then $p_R = x_T - 510, p_G = 255, p_B = 255$,
\end{itemize}
where $x_T$ represents the value of the tile from the bi-variate histogram, while $p_R, p_G$ and $p_B$ are the pixel values, i.e., red, green and blue, respectively. Finally, we observe that if $x_T > 767$, it is clipped to 767---this issue can also be controlled by properly adjusting the chunk size. Instead, for the upper figures, in line with the logic in~\cite{alhazbi2023_ccnc}, we only consider one layer and thus, $0 \le x_T \le 255$.
}

\begin{figure}
    \centering    \includegraphics[width=1.1\columnwidth]{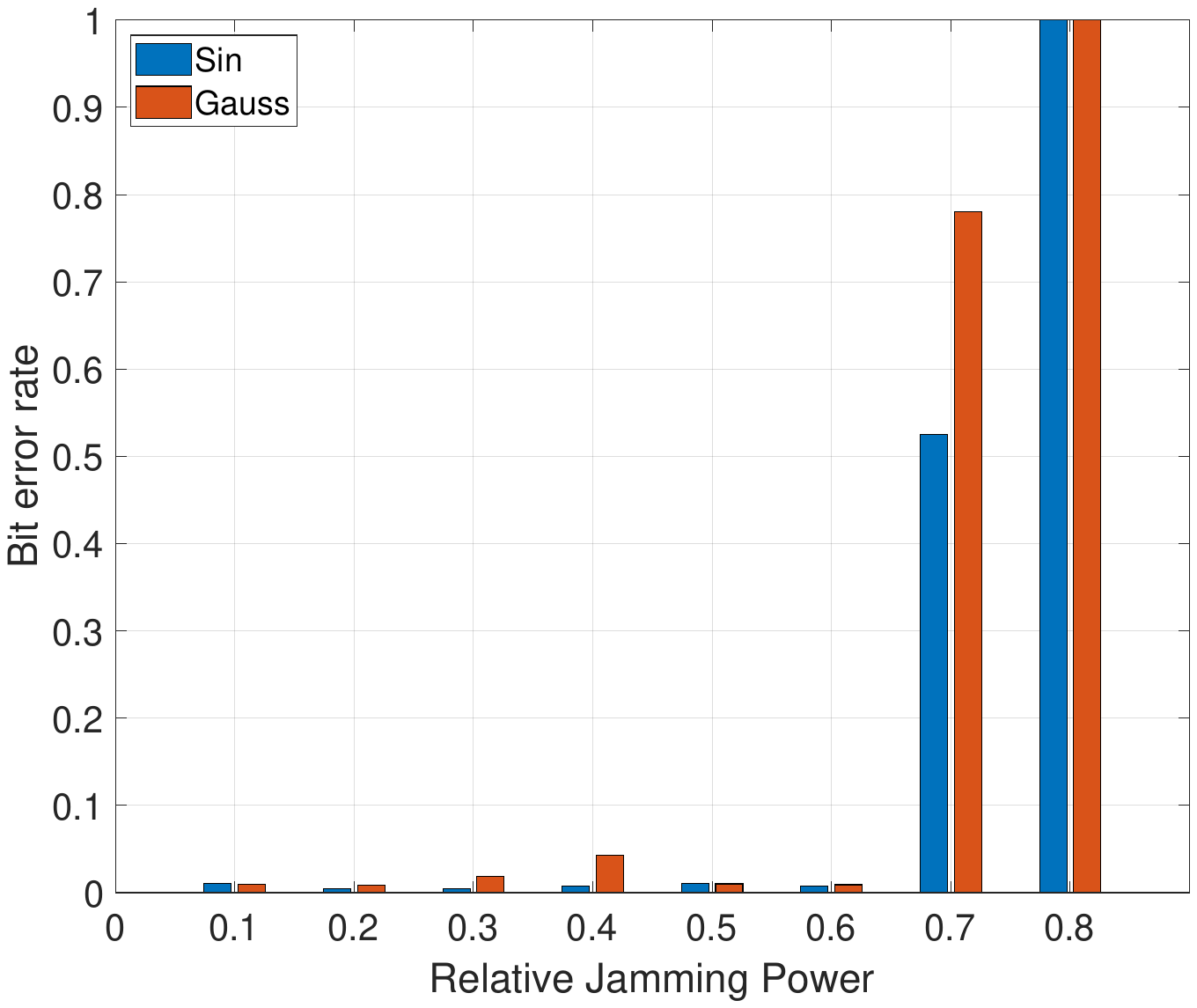}
    \caption{Analysis of the \ac{BER} of the TX-RX communication link using the \acp{SDR} USRP X310, with various levels of RJP and two jamming signals, i.e., tone jamming and Gaussian jamming. We placed the receiver $10$~m away from the jammer.} 
    \label{fig:ber_usrp}
\end{figure}

{\bf Measurements Characterization.} As an introduction to the presentation of our results, with reference to the setup using the USRP X310 \acp{SDR}, in Fig.~\ref{fig:ber_usrp}, we show the \ac{BER} of the TX-RX communication link experienced with different values of the \ac{RJP}, under different jamming signals.

We highlight that when $RJP < 0.7$, the impact of the jammer on the communication link in terms of \ac{BER} is minimal, i.e., very few bits are corrupted. We achieved the lowest BER value with $RJP=0.1$, where $BER\approx1e-6$. On the contrary, most bits are corrupted when $RJP \ge 0.7$. Recall that in this manuscript, we are specifically interested in detecting jamming in a mobile scenario in a low-BER regime, i.e., before its impact on the communication link becomes significant and significantly affects the throughput of the communication link. In this context, we are particularly interested in improving the jamming detection performance when $RJP < 0.7$. For higher values of $RJP$, other techniques based on the analysis of \ac{BER} can already detect jamming \emph{ex-post}, i.e., once the communication link is significantly affected.

Finally, in this context, note that we configured the LimeSDR setup with an absolute transmission gain at the jammer of $23$~dBm. This configuration allows us to have a \ac{BER} of the legitimate communication link of approximately $0.001$, which enables us to study the impact of different configuration parameters of the scenario while matching the conditions of the low-BER regime described above. 

{\bf Performance Metrics.} We compare the performance of the methodologies introduced above mainly concerning their \emph{accuracy}, obtained as $acc = \frac{TP + TN}{TP+FP+FN+TN}$, being $TP$ the number of true positives (i.e., jammed images correctly classified), $TN$ the true negatives (i.e., unjammed images correctly classified), $FP$ the false positives (i.e., unjammed images wrongly classified as jammed ones) and $FN$ the false negatives (i.e., jammed images wrongly classified as unjammed ones). For some of the results shown below, we show the \ac{TPR} and \ac{TNR}, computed as $TPR = \frac{TP}{TP+FP}$ and $TNR = \frac{TN}{TN+FN}$, respectively. We obtain our estimates of accuracy, TPR, and TNR using the cross-validation approach used by the authors of \cite{shvetsova2021}, using 10-fold cross-validation. For each result, we report the mean and the $95\%$ confidence intervals, computed using the tool \emph{tinv} provided by Matlab. \textcolor{black}{Tab.~\ref{tab:experiments} summarizes all the experiments discussed below, together with the relevant parameters.}
\begin{table*}[!ht]
\centering
\caption{\textcolor{black}{Experiments carried out in the manuscript and related parameters tested.}
}
\label{tab:experiments}
\color{black}
\begin{tabular}
{{P{1.5cm}|P{1cm}|P{1.7cm}|P{1.7cm}|P{1.7cm}|P{1.1cm}|P{1.7cm}|P{1.5cm}}}
\textbf{Techniques}                        & \textbf{RJP} & \textbf{Distances [m]}          & \textbf{Samples per Image}              & \textbf{Training Set Size} & \textbf{Jamming Radios} & \textbf{Jamming Oversampling Rates} & \textbf{Jamming Signal Type} \\ \hline
\sol                        & 0.1 - 0.8    & 3, 5, 7, 10, 13, 16, 19, 21 & 10,000 50,000 100,000 500,000 1,000,000 & 2, 9, 18, 36, 54, 72       & 4, 5, 6, 7              & 1, 2, 3, 4                          & AWGN, BPSK                   \\
\cite{alhazbi2023_ccnc}  & 0.1 - 0.8    & 3, 5, 7, 10, 13, 16, 19, 21 & 10,000 50,000 100,000 500,000 1,000,000 &                            &                         & 1, 2, 3, 4                          & AWGN, BPSK                   \\
\cite{alhazbi2023_arxiv} & 0.1 - 0.8    & 3, 5, 7, 10, 13, 16, 19, 21 & 10,000 50,000 100,000 500,000 1,000,000 &                            &                         &                           &                  
\end{tabular}
\end{table*}

\subsection{Jamming Detection Robustness in Low-BER Regime}
\label{sec:autoenc_robustness}

\begin{figure}
    \centering    
    \includegraphics[width=\columnwidth]{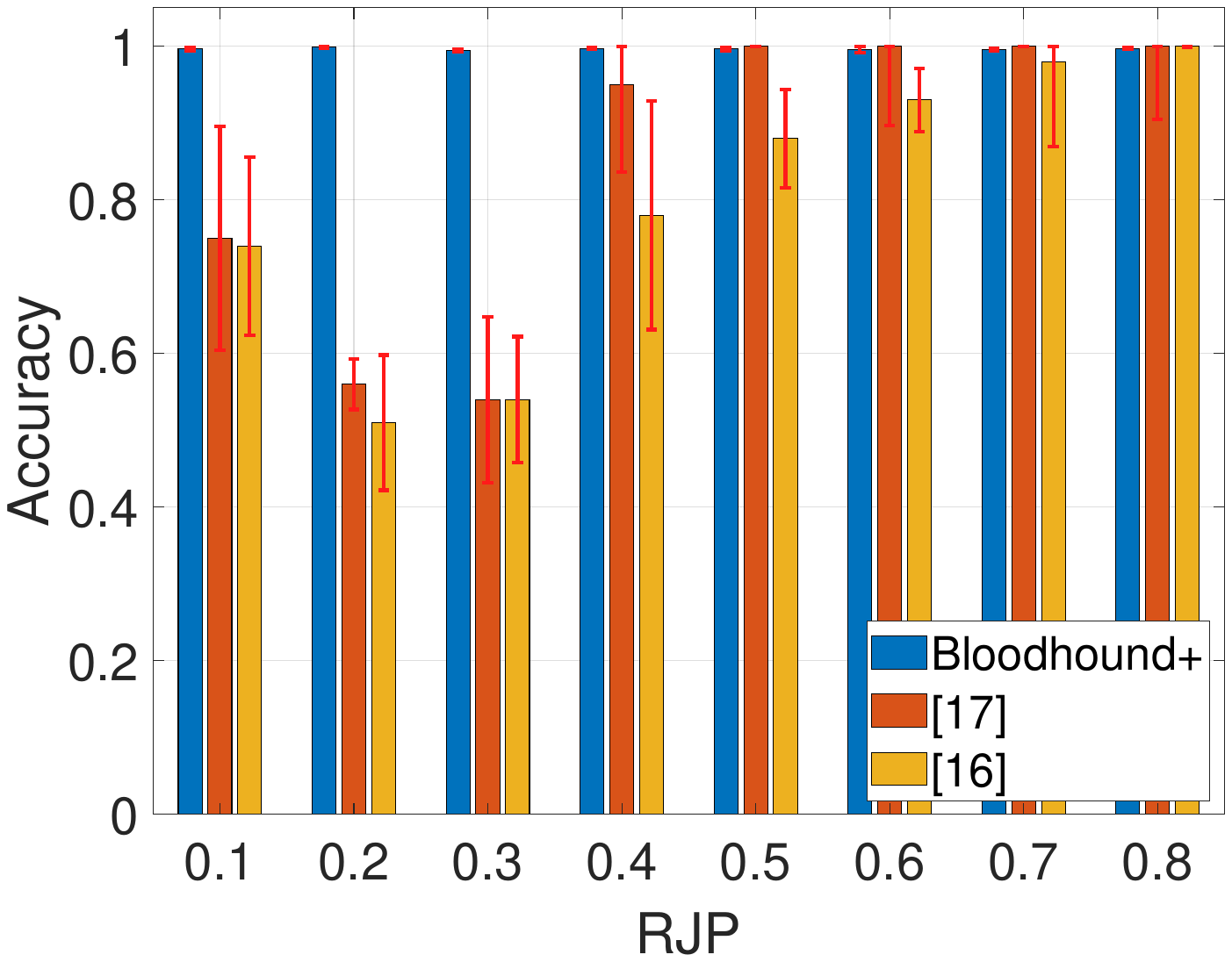}
    \caption{Classification accuracy of \sol, the proposal in~\cite{alhazbi2023_arxiv} and the solution in~\cite{alhazbi2023_ccnc} with various levels of \ac{RJP}. } 
    \label{fig:rjp}
\end{figure}

We first consider the impact of the received jamming power at the receiver on the capability of \sol\ to detect the presence of the jammer. We consider the setup using the Ettus X310 \ac{SDR}, and precisely, the measurements where we placed the receiver $10$~meters away from the jammer. Here, we generate images using $n=10^{5}$ I-Q samples and evaluate the performance of \sol, the proposal in~\cite{alhazbi2023_arxiv} and the solution in~\cite{alhazbi2023_ccnc} to detect jamming, in terms of overall classification accuracy. We report the results of our investigation in Fig.~\ref{fig:rjp}.

First, we note that the approaches in~\cite{alhazbi2023_arxiv} and~\cite{alhazbi2023_ccnc}, based on \acp{CNN}, reliably identify jamming only when $RJP \ge 0.4$. With lower values of the RJP, their performances do not follow a unique trend, and the results also exhibit high variance (see the red bars indicating the confidence intervals of the measurements). On the contrary, \sol\ reports remarkable performances for every value tested of $RJP$, with a minimum accuracy of $0.997$. We believe such a result is due to the rationale of autoencoders used in \sol, which work only on \emph{unjammed images}. This configuration allows auto-encoders to build a profile of the regular behavior of the wireless channel to identify more minor differences (anomalies) reliably. We also note that the performances of \sol\ do not depend on the speed of the involved entities nor the smoothness of the change of channel conditions. Indeed, our approach processes chunks of $n$ I-Q samples and compares the images created from such samples with the expected channel conditions acquired during training time.

\begin{figure}
    \centering    
    \includegraphics[width=\columnwidth]{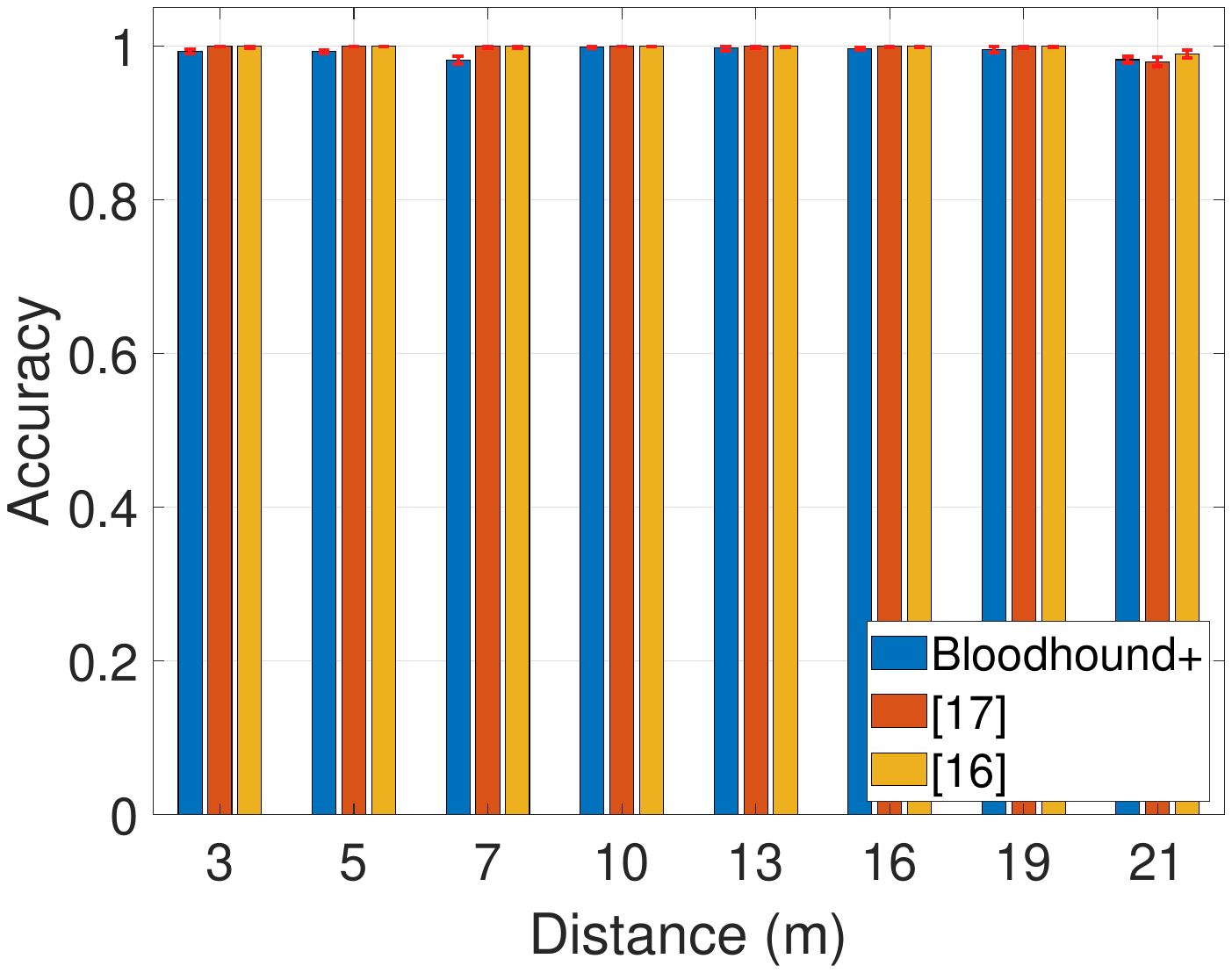}
    \caption{Classification accuracy of \sol, the proposal in~\cite{alhazbi2023_arxiv} and the solution in~\cite{alhazbi2023_ccnc} when positioning the receiver at various distances from the jammer. } 
    \label{fig:dist}
\end{figure}

We also investigated the impact of the receiver's distance from the jammer on the performance of the cited solutions. To this end, using the same setup used for the previous tests, we set $RJP=0.5$ and move the receiver away from the jammer to a distance of $21$~meters. We stopped at such a distance due to the physical limitations of the involved hardware, i.e., at distances higher than $21$~meters, the \ac{BER} of the legitimate communication link increases significantly, preventing us from performing the test reliably. 
\textcolor{black}{We first acknowledge that the receiver experiences increasing interference while moving closer to the jammer. This phenomenon is shown in Fig.~\ref{fig:rss_dist}, highlighting that the process is not linear but follows a polynomial model while increasing distance, i.e., $a x^2 + bx + c$. Note that our result is consistent with the findings of Tedeschi et al. reported in~\cite{tedeschi2021_spaccs}.
\begin{figure}[h]
  \centering
  \includegraphics[angle=0, width=\columnwidth]{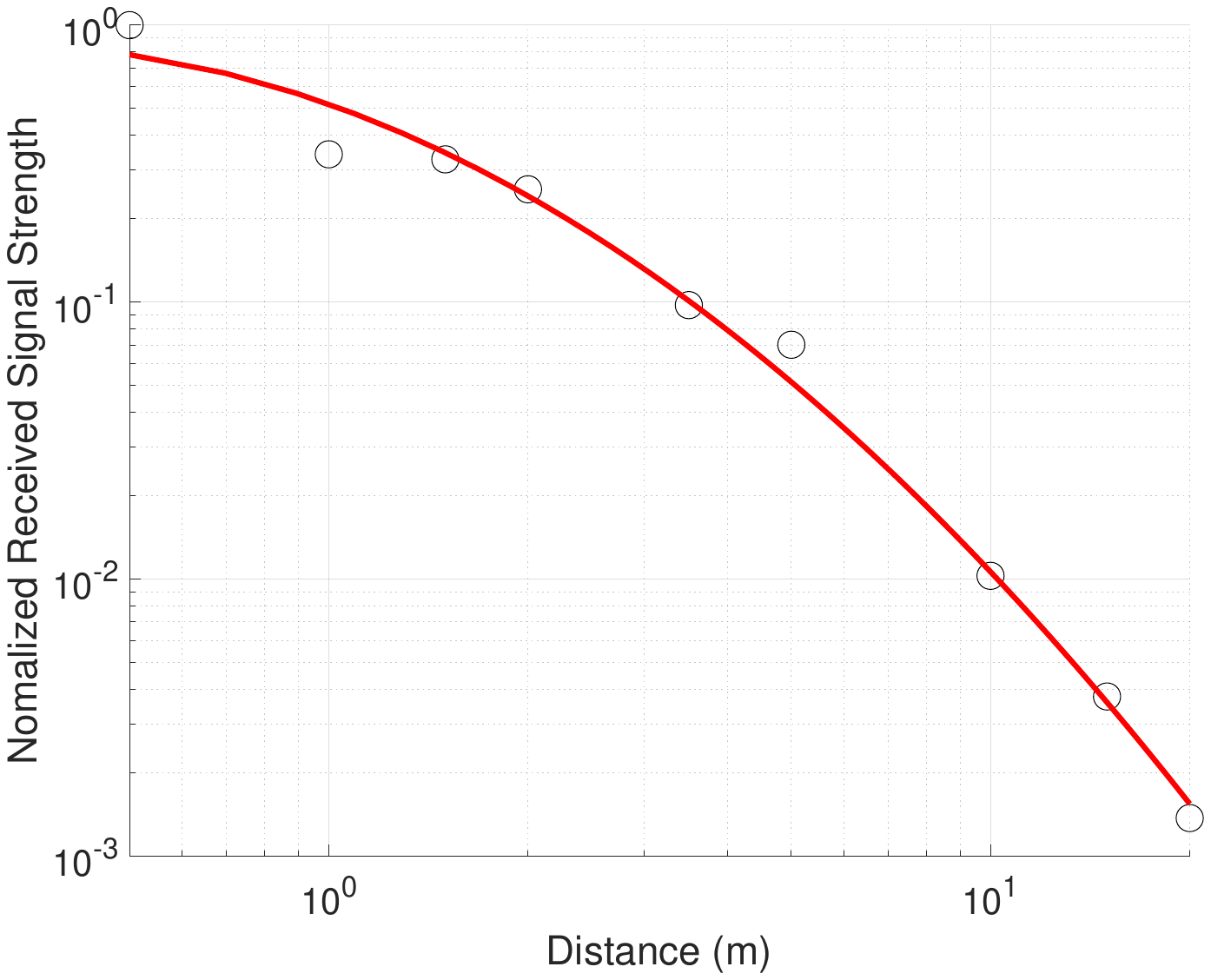}
  \caption{\textcolor{black}{Normalized Received Signal Strength (NRSS) as a function of the distance: the NRSS decreases when the receiver moves far away from the transmitter. The solid red line shows the model that best fits the experimental data (black circles).}}
  \label{fig:rss_dist}
\end{figure}
We also notice that, for specific distance values, significantly different RSS values may be experienced, due to the fast-fading process affecting wireless communication channels. However, recall that lower RSS does not necessarily imply higher BER. As our manuscript focuses on jamming detection in low-BER scenarios, our analysis considers a condition where the BER of the communication channel is low to provide jamming detection before the loss of the communication link.}
We report in Fig.~\ref{fig:dist} the results of our experiments involving jamming detection via \sol\ and competing approaches.

Here, we notice that all tested solutions report remarkable performance with minimal differences. All three approaches reliably detect jamming at various distances, with an average accuracy well above $0.99$. Overall, the distance between the receiver and the transmitters does not affect the classification accuracy, which remains very high even when the receiver is $21$~meters away. We present such a result primarily to show that, similarly to competing approaches, \sol\ can detect jamming even at a significant distance from the jamming source. In addition to such a result, we present additional results below that show the enhanced robustness of \sol\ compared to other approaches.   

\begin{figure}
    \centering    
    \includegraphics[width=\columnwidth]{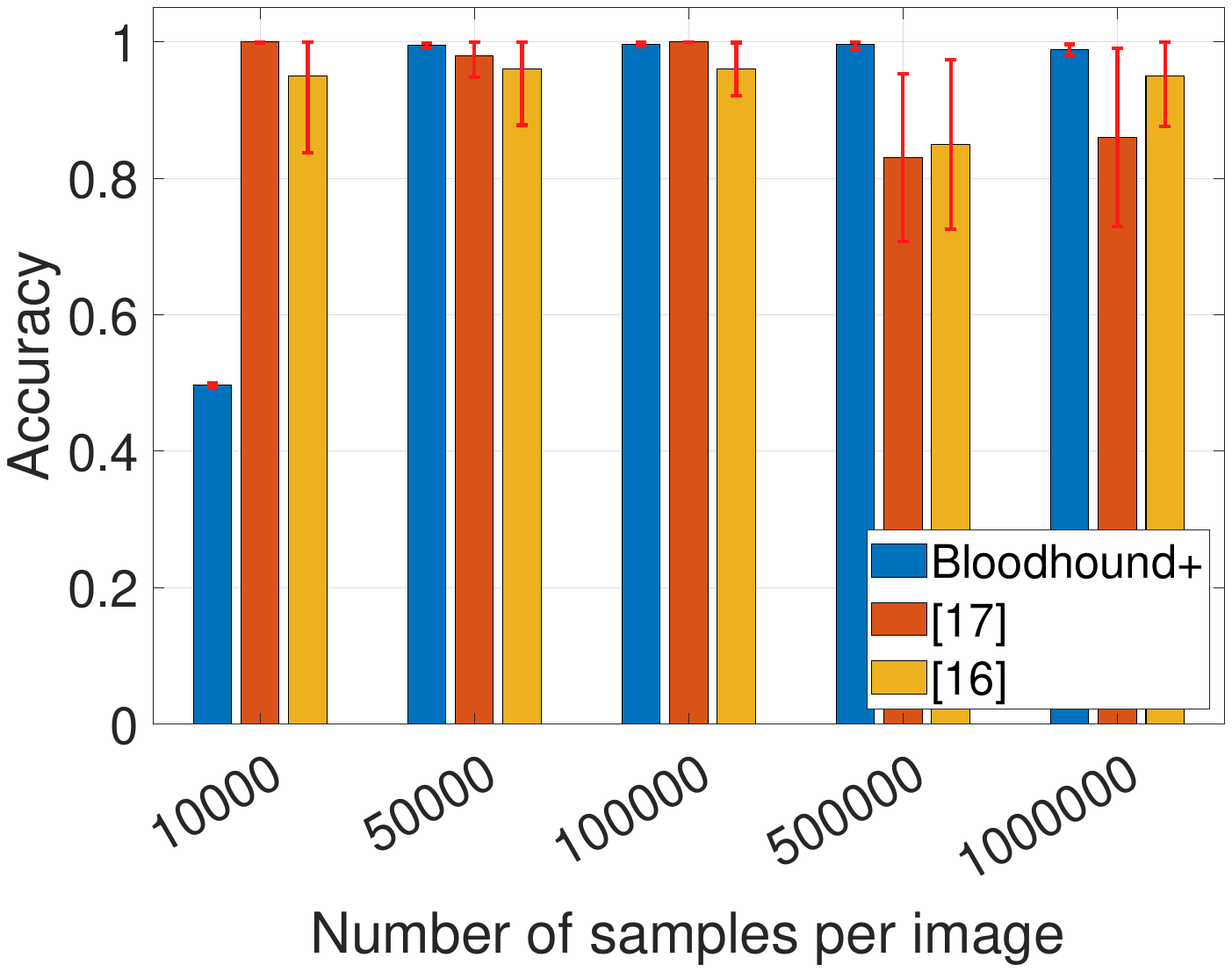}
    \caption{Classification accuracy of \sol, the proposal in~\cite{alhazbi2023_arxiv} and the solution in~\cite{alhazbi2023_ccnc} considering an increasing number of samples $N$ in the \emph{Image Generation} process. } 
    \label{fig:samplesImage}
\end{figure}

Another critical parameter of \sol\ is the number of samples used to generate images, namely $n$. The higher the value of this parameter, the higher the number of samples to use for image generation. Thus, the longer the receiver has to acquire samples from the wireless channel, the higher the processing overhead of the solution. For this analysis, in the same setup as in the previous experiments, we considered the data acquired with the receiver located $10$~m away from the jammer and $RJP=0.5$, and tested the performance of the three approaches while increasing the number of samples used in the image generation phase, from $10,000$ to $1,000,000$. Fig.~\ref{fig:samplesImage} reports the results of our analysis.

Note that the solutions in~\cite{alhazbi2023_arxiv} and~\cite{alhazbi2023_ccnc} report a higher classification accuracy than \sol\ for a low number of samples. For example, when the $n=10,000$, \sol\ reports an accuracy of $0.498$ while such values amount to $0.99$ and $0.95$ for the solution in~\cite{alhazbi2023_arxiv} and~\cite{alhazbi2023_ccnc}, respectively. When the available number of samples increases, the accuracy of the benchmark approaches is still high, but the variance becomes larger (see the red bars). On the contrary, when $n \ge 50,000$, not only the accuracy of \sol\ is very high (always higher than $0.99$), but the variance is also minimal (less than $0.001$ for all tests), indicating greater robustness and reliability. Such results further motivate the deployment of \sol\ and highlight its superiority compared to the benchmark solutions.

\begin{figure}
    \centering    
    \includegraphics[width=\columnwidth]{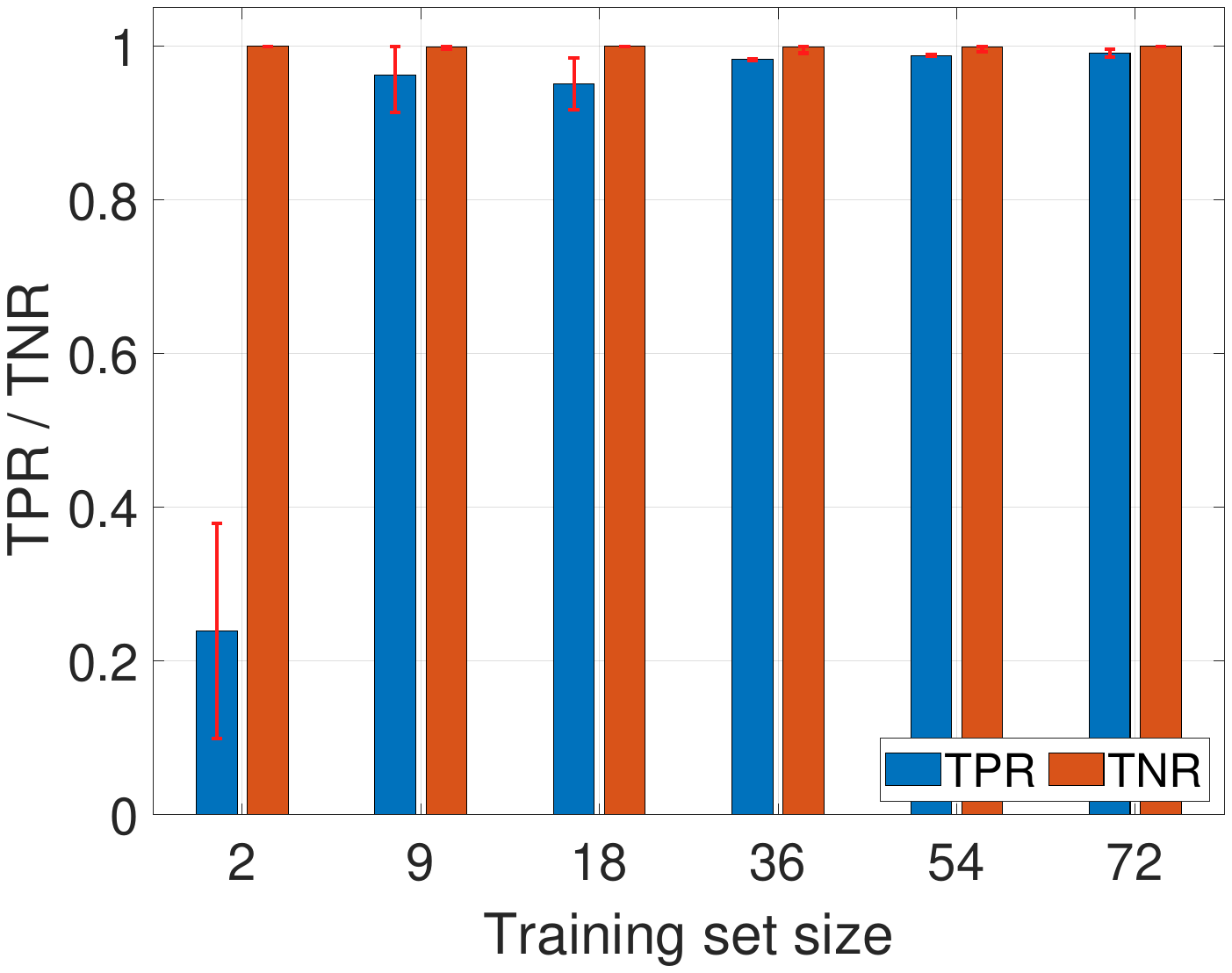}
    \caption{\ac{TPR} and \ac{TNR} of \sol\ when varying the training set size, i.e., the number of \emph{unjammed} images used for training the autoencoder. } 
    \label{fig:setSize_autoenc}
\end{figure}

To provide further insight into the performance of \sol, we investigated the impact of additional configuration parameters. In particular, in the same setup as the last cited test, considering $n=100,000$, we evaluated the effect of the training set size. Figure~\ref{fig:setSize_autoenc} summarizes the results of our investigation, distinguishing the achieved \ac{TPR} and \ac{TNR} of our proposed solution.

To perform reliably, \sol\ requires a minimum training set of only $9$ images, reporting $TPR$ and $TNR$ values of $0.962$ and $0.99$, respectively. Performance remains almost constant when increasing the training set size. However, increasing the training set could be particularly relevant in very noisy scenarios characterized by a broader range of \emph{expected} wireless channel fluctuations. We recall that we need to train \sol\ only once before deployment, and such results do not affect the deployability of our solution at runtime.

\begin{figure}
    \centering    
    \includegraphics[width=\columnwidth]{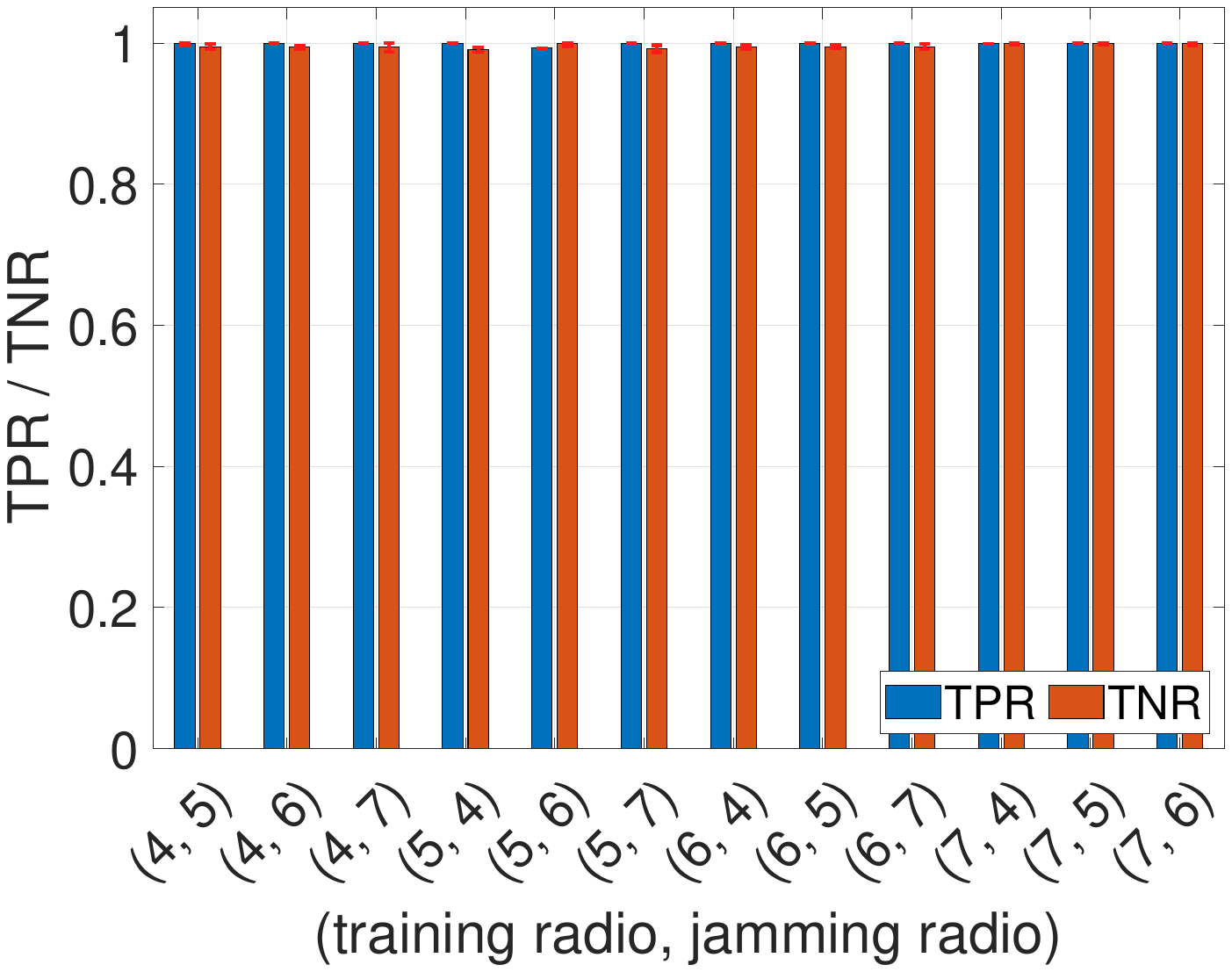}
    \caption{\ac{TPR} and \ac{TNR} of \sol\ when varying the hardware used for jamming and the radio used for training. } 
    \label{fig:hw_autoenc}
\end{figure}

We also investigated further any bias of our results concerning the specific hardware used for the experiments. Taking into account the same scenario as in previous experiments, i.e., the receiver located $10$ meters away from the jammer, we evaluated the \ac{TPR} and \ac{TNR} of \sol\ when changing the hardware used for jamming among the five available radios. We highlight that this methodology prevents the autoencoder from fingerprinting both the transmitter and the receiver, these being the same for all measurement classes. 
Moreover, we considered different hardware for the jammer during our measurements, i.e., we mutually excluded the ones adopted for training from the ones adopted for testing. The mentioned strategy eventually guarantees that the autoencoder learns the characteristics of the legitimate signal only while being independent of the transmitter, the receiver, and the jammer hardware (we do not use jammed signals for training).
Specifically, we consider all unjammed images obtained when placing the receiver at a distance of 10 meters from the transmitter, using the radio $x$ as the jammer. Next, we consider all the jammed images generated with the receiver located at a distance of 10 meters and use radio $y$ as a jammer. We separated the unjammed and jammed images into 10 folds, and we trained \sol\ on 9 of the folds containing unjammed images, holding the $i$-th one to estimate the TNR. Next, we estimate the TNR on the $i$-th fold of the unjammed images and evaluate the TPR on the $i$-th fold of the jammed images. Fig.~\ref{fig:hw_autoenc} reports the result of our analysis. The tick labels on the x-axis are of the form $(x,y)$, as described above.

Note that the TPR and TNR remain almost unchanged while varying the hardware used for jamming and the radio considered for training. Therefore, we can safely assume that \sol\ is not biased by the specific radios used in the experiments. Still, it can extract the features of the wireless channel useful for detecting jamming independently of the particular hardware. 

\subsection{Impact of Different Hardware and Sampling Rate}
\label{sec:samplingRate}

\begin{figure}
    \centering    
    \includegraphics[width=\columnwidth]{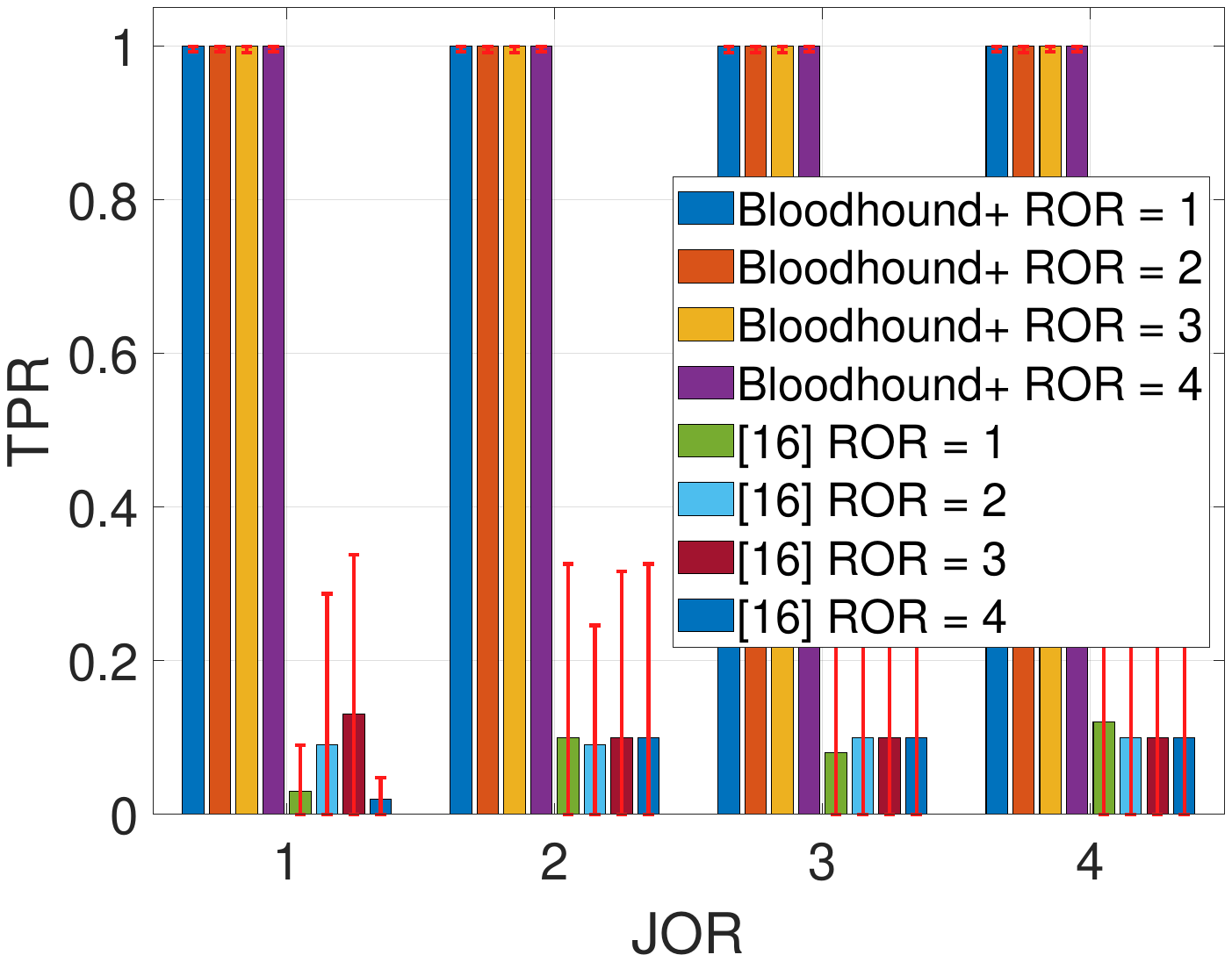}
    \caption{ True Positive Ratio (TPR) of \sol\ and the proposal in~\cite{alhazbi2023_ccnc}, with various \ac{ROR} and \ac{JOR} values, when the jammer injects Gaussian noise (Gaussian random jamming). } 
    \label{fig:randomJam}
\end{figure}

We obtained all the results shown in the previous subsection using a single hardware brand (Ettus Research USRP X310) and a single configuration of the sample rate, i.e., $1$~Msa/s. 
To further extend our experimental assessment, we first analyzed the data collected through the second setup described in Section~\ref{sec:measurements}, adopting the hardware LimeSDR Mini. An important consideration about the scenario described in Section~\ref{sec:sys_adv_model} is that the jammer might not know the adopted sampling rate in advance. Thus, to disrupt ongoing communications as much as possible, in real-life scenarios, the jammer might emit jamming using the maximum achievable sampling ratio, likely higher than the one adopted by the legitimate communication link. At the same time, the receiver might oversample the signal, obtaining more helpful information for jamming detection. This additional information might be discarded for communication but might benefit jamming detection. Therefore, in our tests, we trained 
\sol\ on unjammed images of the legitimate communication link obtained under $RJP=0.5$ and a distance of 3 meters. Then, we tested using unjammed images (disjoint from those used for training) and jammed images obtained from I-Q samples generated with various levels of \ac{JOR}. For this test, we let the jammer emit random Gaussian noise. We also compared the performance of \sol\ with the solution in~\cite{alhazbi2023_ccnc}. Fig.~\ref{fig:randomJam} summarizes the results of our analysis in terms of \ac{TPR} (for comparison purposes).
Note that the TNR of \sol\ was also excellent, at an average of 0.987 over all values of JOR. 

We can distinguish two effects from the reported results. Considering the configuration with $ROR=1$ and $JOR=1$, we notice that the solution in~\cite{alhazbi2023_ccnc} already reports very low TPR ($0.03$). We highlight that this result is not related to different oversampling rates but to the new hardware used for the experiments. Indeed, the LimeSDR is a cheaper hardware, which introduces additional inaccuracies and imperfections in the I-Q samples received. Such inaccuracies affect the shape of the I-Q samples, which is now more spread around the expected symbol than before, leading to jammed images different from the ones in the training set of the solution in~\cite{alhazbi2023_ccnc}. 
The approach in~\cite{alhazbi2023_ccnc} does not catch these variations, thus failing to perform reliable jamming detection. On the contrary, in such a configuration, the performance of \sol\ does not change compared to the results shown in Section~\ref{sec:autoenc_robustness}, demonstrating once again the enhanced robustness offered by the autoencoders used in \sol\ compared to the \acp{CNN} used in the competing solution.
With higher ROR and JOR values, the performance of the approach in~\cite{alhazbi2023_ccnc} remains well below $0.2$, confirming the unsuitability of such a solution for detecting jammers in the wild. Instead, \sol\ can mitigate and overcome the impact on the wireless channel of different oversampling ratios, being able to detect jamming also when the \ac{JOR} is very high. 

\subsection{Deceptive Jamming}
\label{sec:deceptive}

In previous experiments, we mainly considered two jamming models, that is, tone jamming (using a sinusoid signal) and random jamming (using \ac{AWGN}). When the adversary does not know the modulation used by the legitimate communication link, the jamming models mentioned and investigated above are the most reasonable options to disrupt the channel as much as possible.
However, more powerful attackers might know in advance or become aware at runtime (e.g., by eavesdropping) of the modulation used by the legitimate communication link. Based on this knowledge, they might use the signal as part of a jamming attack optimized to disrupt the communication link as much as possible. Recall that the legitimate communication link in our experiments adopts the \ac{BPSK} modulation scheme. In such a scenario, the usage of a jamming signal modulated also as a \ac{BPSK} allows to boost the effectiveness of the jamming activity, as acknowledged by several scientific contributions both analytically~\cite{amuru2015_tifs} and experimentally~\cite{punal2014_wowmom}.

\begin{figure}
    \centering    
    \includegraphics[width=\columnwidth]{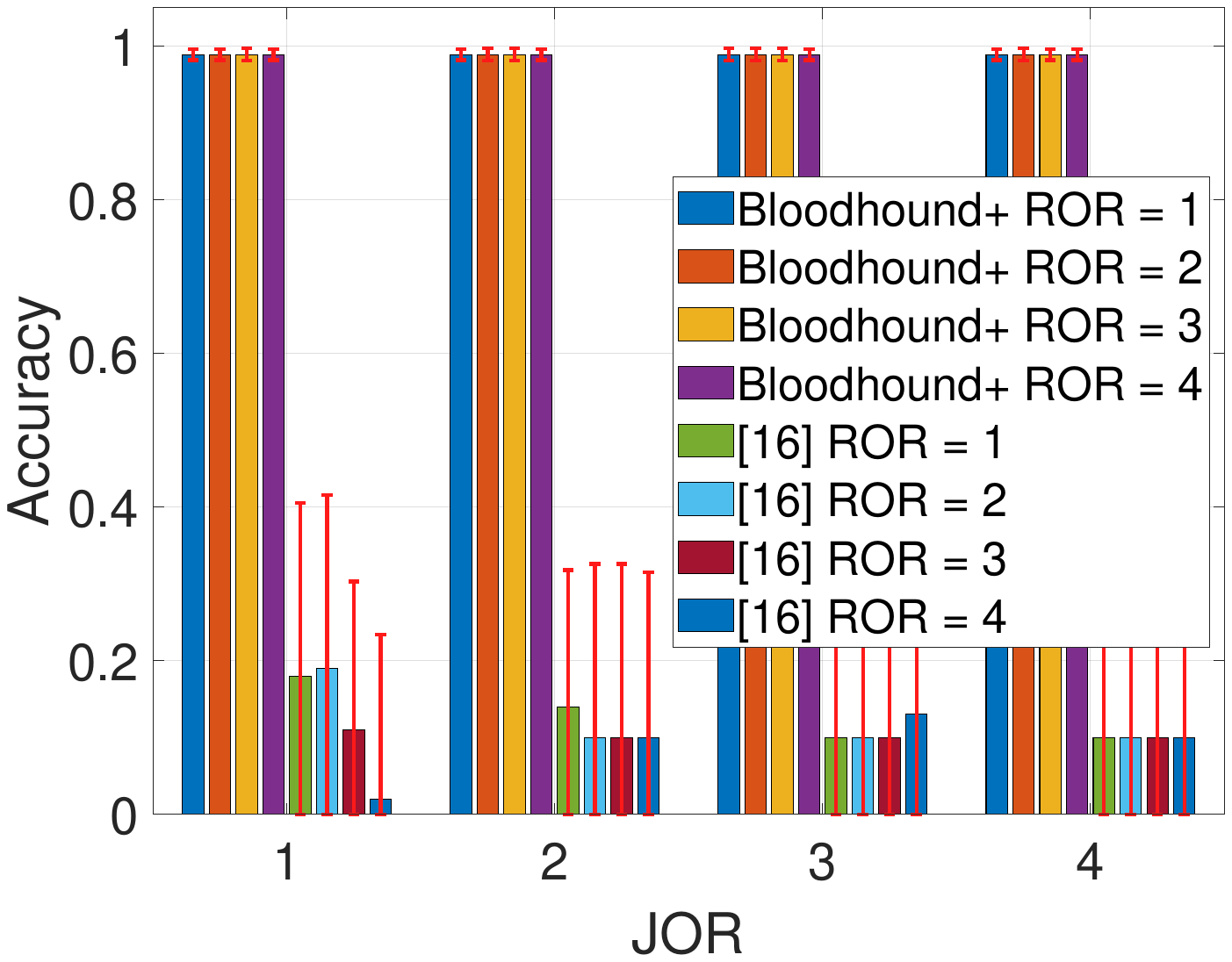}
    \caption{True Positive Ratio (TPR) of \sol\ and the proposal in~\cite{alhazbi2023_ccnc}, with various \ac{ROR} and \ac{JOR} values, when the jammer injects the same signal from the legitimate communication link (deceptive jamming). } 
    \label{fig:deceptiveJam}
\end{figure}

To this end, using the same setup as the previous experiments in Section~\ref{sec:samplingRate}, we evaluated the ability of both \sol\ and the solution in~\cite{alhazbi2023_ccnc} to detect \emph{deceptive jamming}, i.e., a jammer injecting the same signal used as part of the legitimate communication link. Fig.~\ref{fig:deceptiveJam} summarizes the results of our analysis in terms of TPR (for comparison purposes). Note that also in this case, the TNR of \sol\ is 0.987, on average over all JOR, since we trained the autoencoder in \sol\ with the same data as in Fig.~\ref{fig:randomJam}. 

Even when the adversary uses deceptive jamming, \sol\ significantly outperforms the benchmark solution in all tested configurations, showing perfect TPR and being robust to adopting a high \ac{JOR} by the adversary. 

Overall, the results reported above demonstrate the superiority of \sol\ compared to benchmark approaches and the robustness of our solution to a wide range of configuration parameters and scenarios, making it the preferred solution for jamming detection in a low-BER regime.

\textcolor{black}{
\subsection{Considerations on Interference}
\label{sec:interference}
Various factors, such as interference and obstacles, might affect the performance of \sol. We show in Fig.~\ref{fig:multipath} the effect on the I-Q samples of people passing through our experimental setup.
\begin{figure*}
\centering
\begin{minipage}{.19\textwidth}
  \centering
  \includegraphics[width=\linewidth]{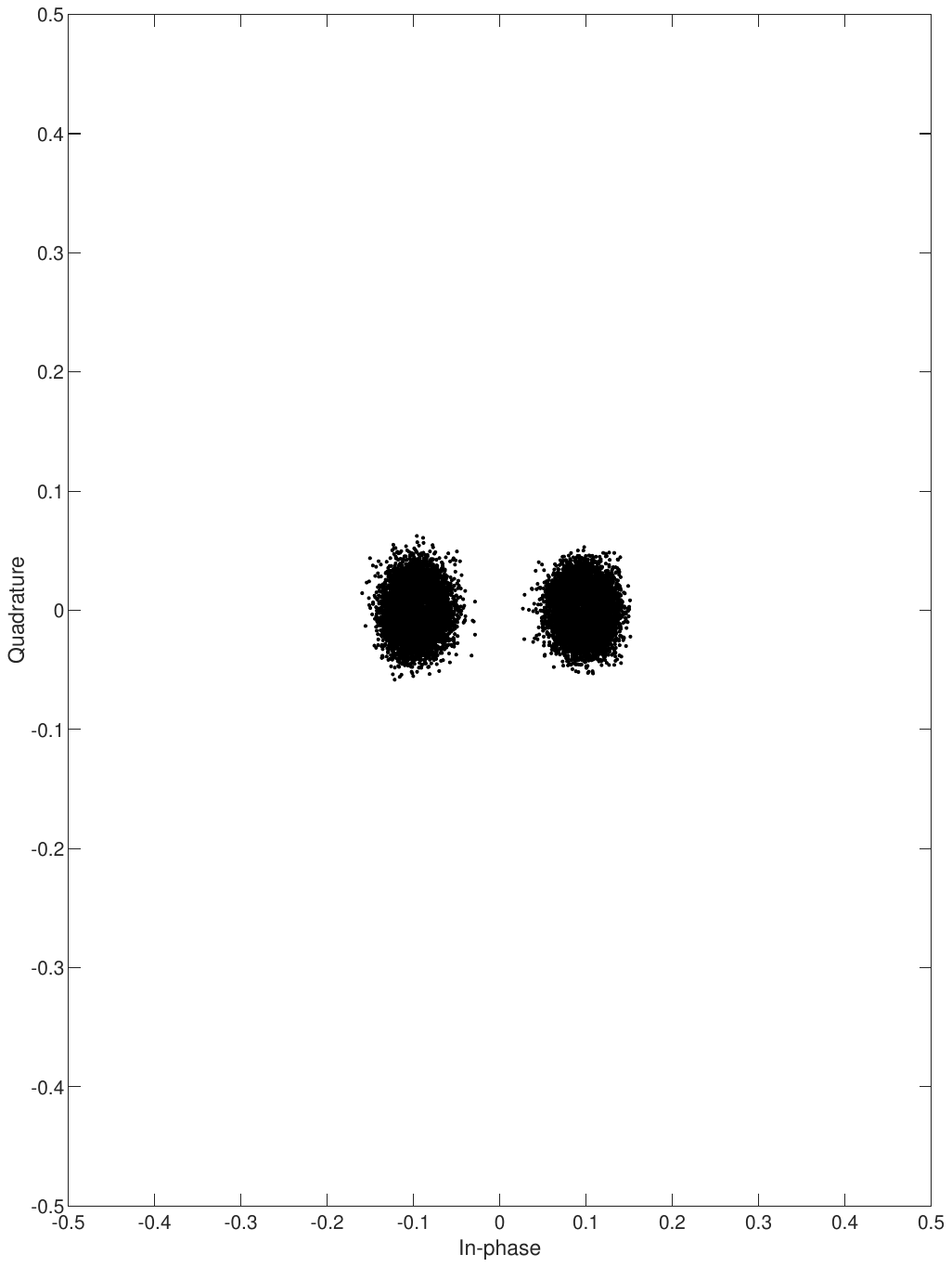}
    \subcaption{t = 0}
\end{minipage}
\begin{minipage}{.19\textwidth}
  \centering
  \includegraphics[width=\linewidth]{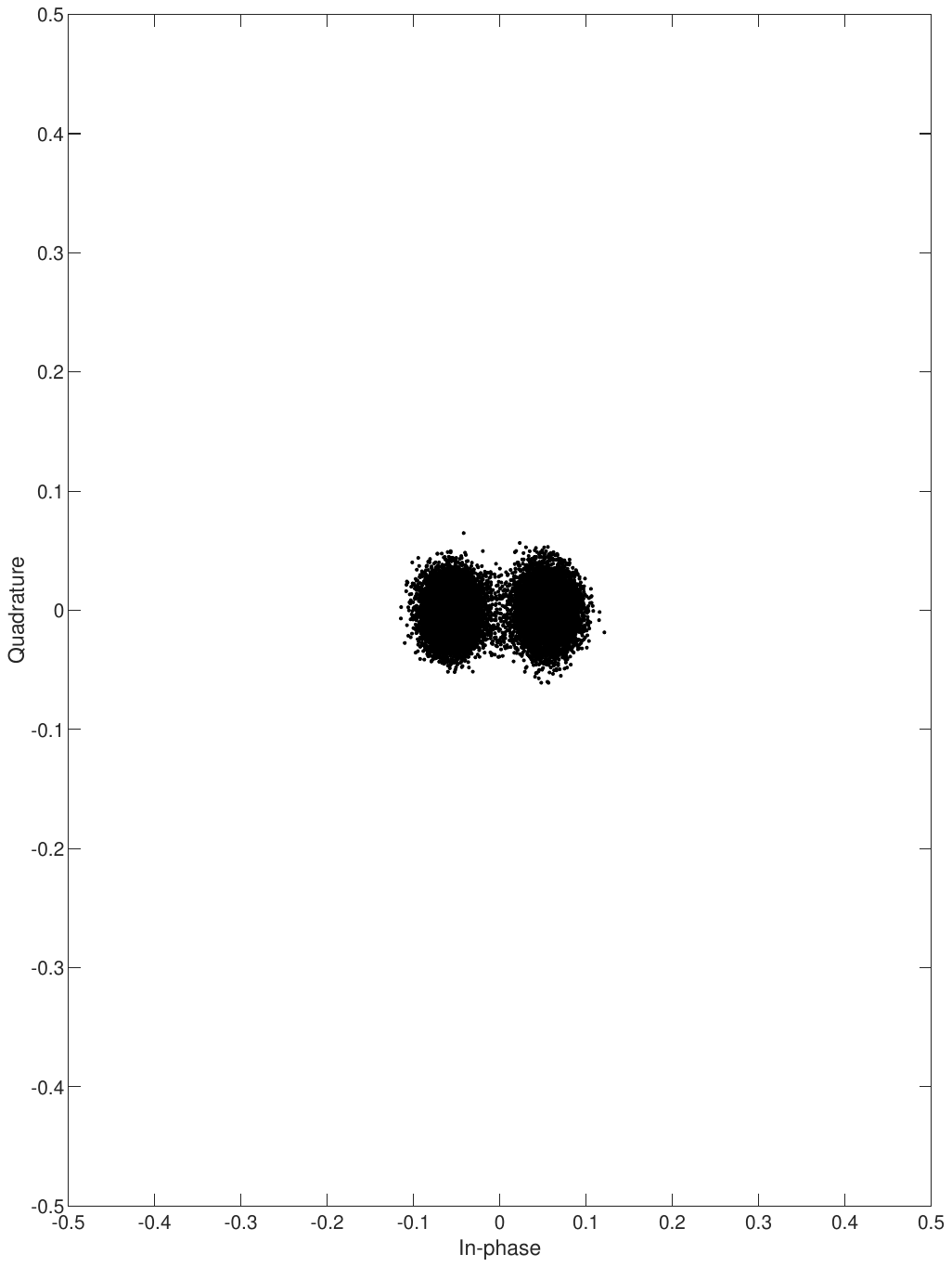}
    \subcaption{t = 1}
\end{minipage}
\begin{minipage}{.19\textwidth}
  \centering
  \includegraphics[width=\linewidth]{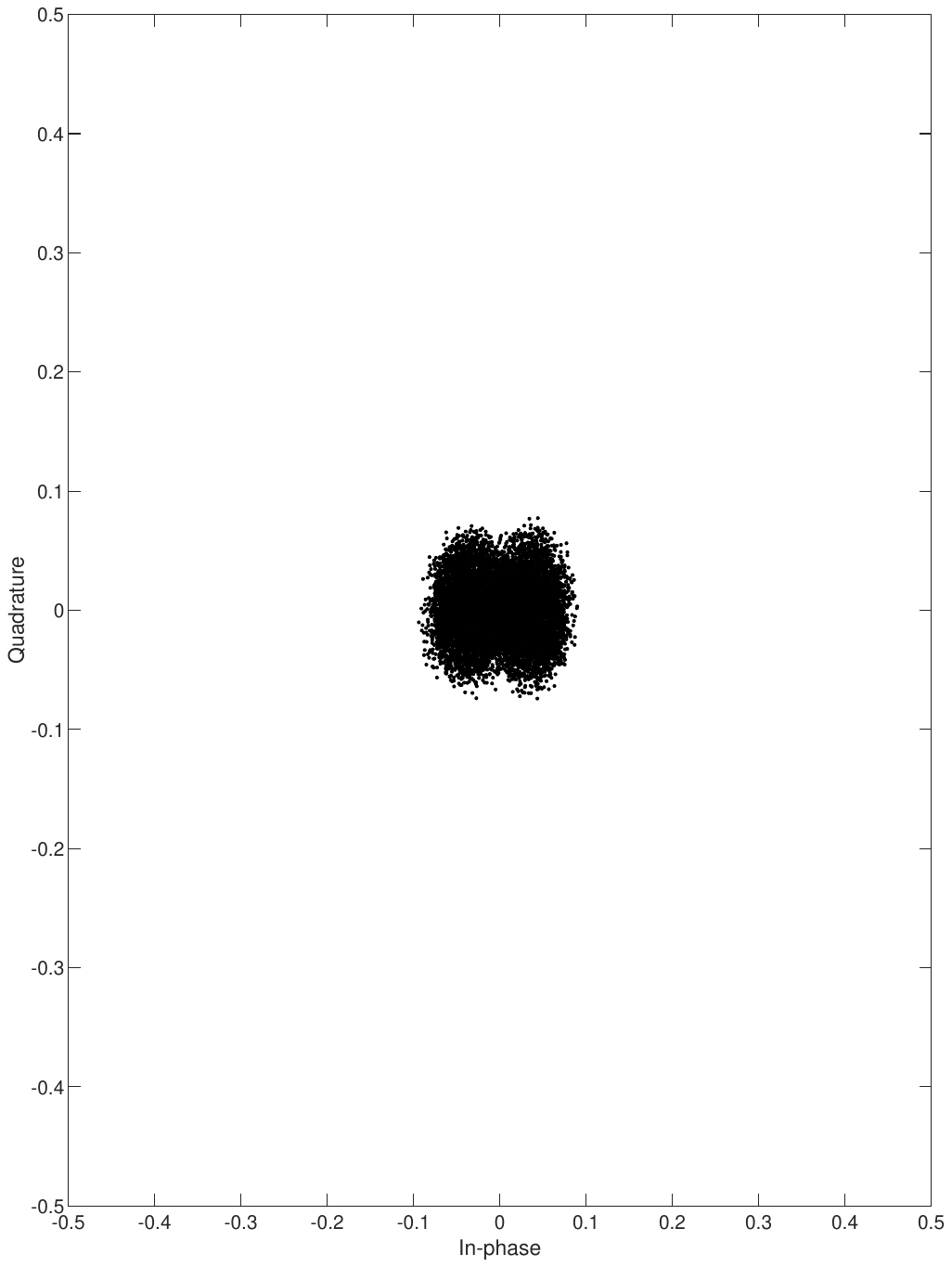}
    \subcaption{t = 2}  
\end{minipage}
\begin{minipage}{.19\textwidth}
  \centering
  \includegraphics[width=\linewidth]{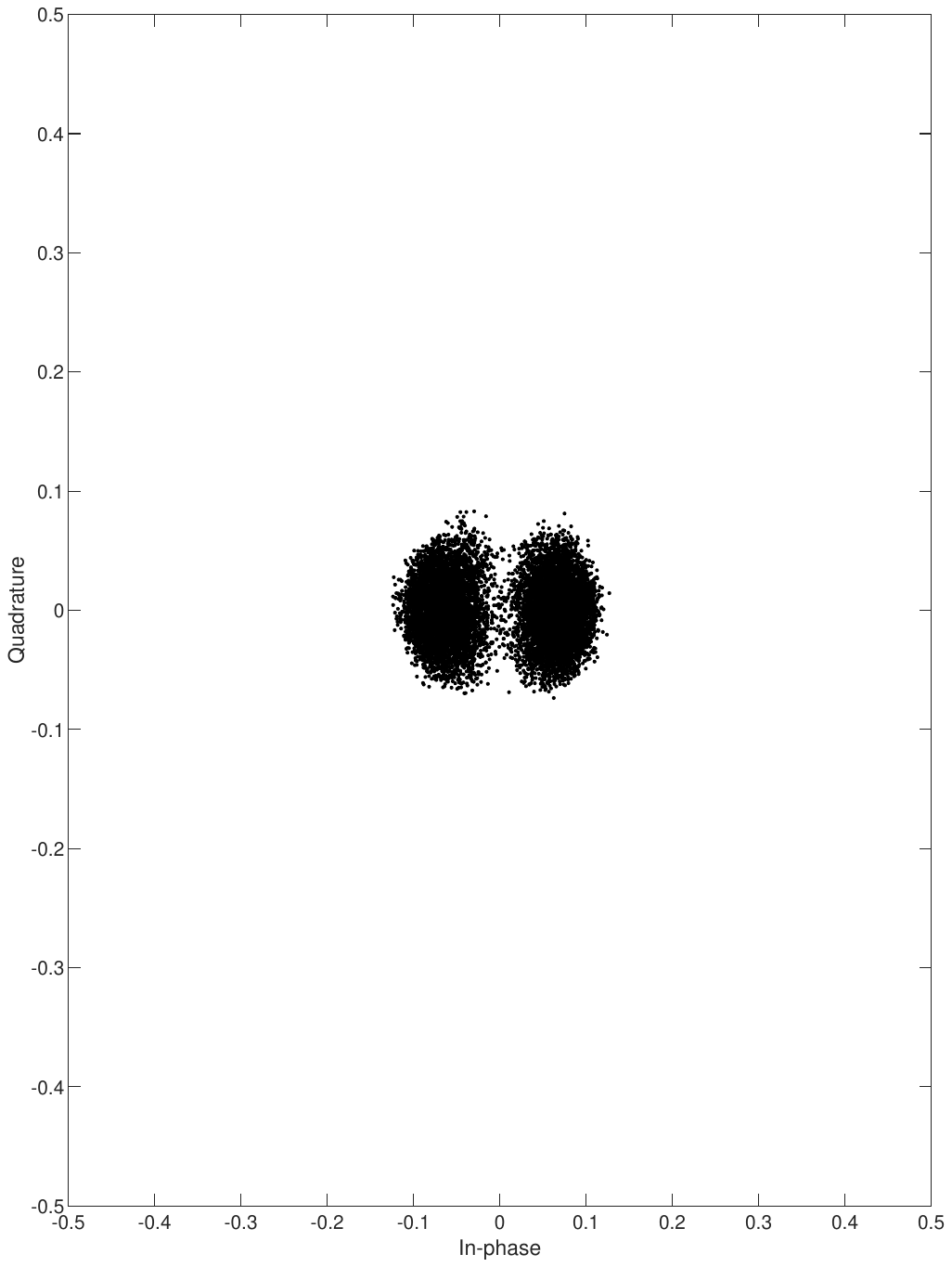}
    \subcaption{t = 3}
\end{minipage}
\begin{minipage}{.19\textwidth}
  \centering
  \includegraphics[width=\linewidth]{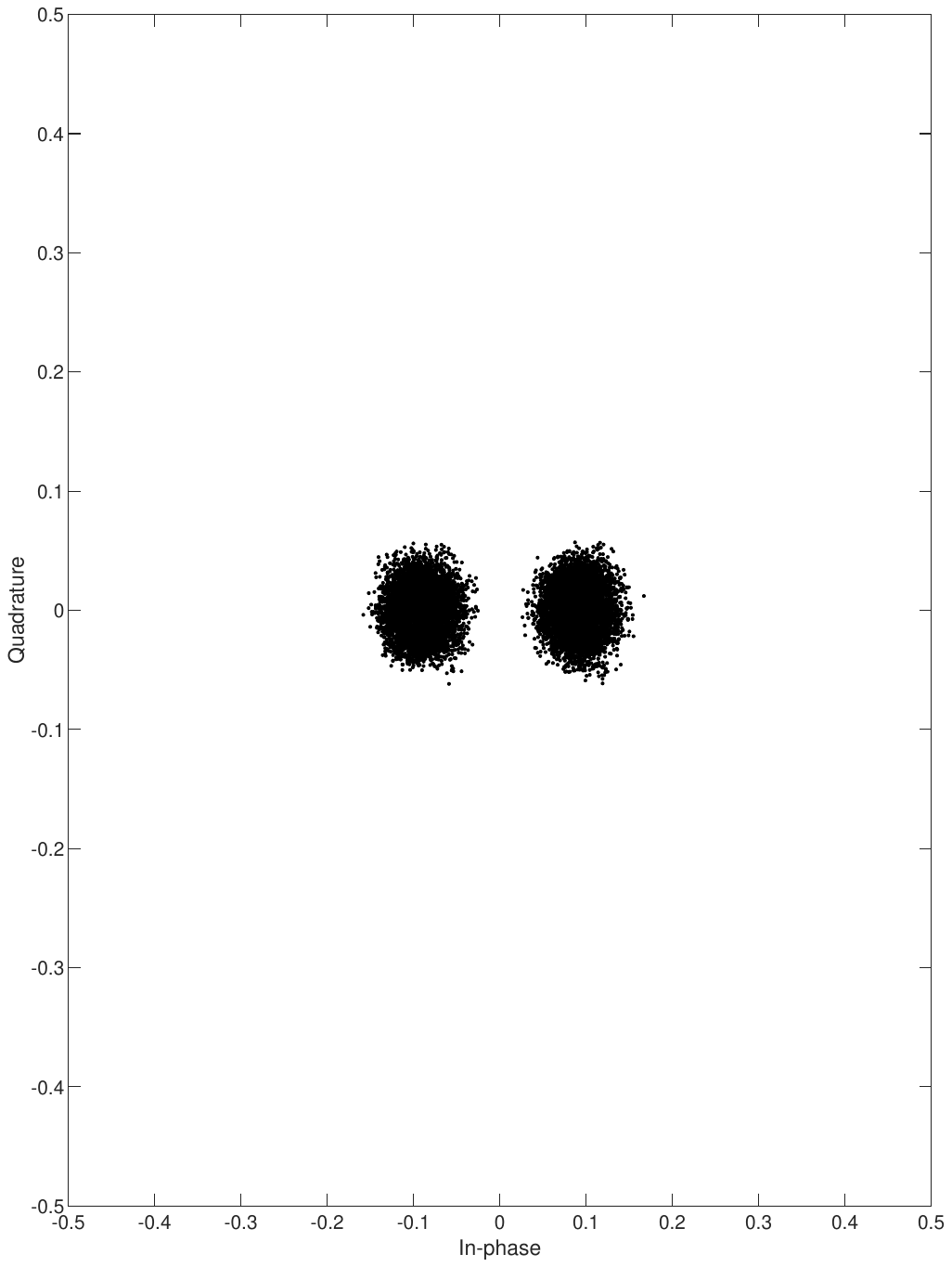}
    \subcaption{t = 4}
\end{minipage}
\caption{\textcolor{black}{
The effect of multipath on the transmitter-receiver link across five time windows. Compared to an ideal (static) scenario (Figs. (a) and (e)), the shape of the I-Q samples at the receiver (modulated through the BPSK scheme) is significantly affected (see Figs. (b), (c), and (d)) when an event is happening, e.g., moving objects.}
}
\label{fig:multipath}
\end{figure*}
We notice that the shape of the I-Q samples at the receiver is significantly affected compared to a LOS scenario. Also, compared to jamming, note that these phenomena are much quicker, and the shape of the I-Q samples returns to the regular shape much faster than under jamming attacks. Thus, to allow \sol\ not to declare jamming in such cases, we need to experience such phenomena during the training phase to make them part of the regular conditions of the communication channel, especially indoors. These considerations motivate the deployment of our solution and the experiments in an office environment during working hours: we experience several instances of such phenomena during training, and they contribute to creating an expected profile of unjammed scenarios which takes interferences into account. During jamming attacks, the shape of the I-Q samples is displaced from the expected profile for a much extended time, contributing to enhancing the performance of our solution.
}


\section{Conclusion}
\label{sec:conclusion}

In this paper, we have presented \sol, an approach that allows drones and possibly other mobile devices to detect jamming at the physical layer (PHY) of the communication stack. Our solution works on raw I-Q samples extracted from the communication link, converts them into grayscale images, and uses \emph{sparse autoencoders} to detect discrepancies with the expected profile of the channel. Therefore, \sol\ can detect jamming in low-BER regimes, i.e., well before its effect could cause a significant decrease in the quality of the main communication link. At the same time, our solution allows drones to efficiently avoid the jammed area and maintain complete control and safety.
To test the effectiveness of our solution, we conducted an extensive measurement campaign, acquiring real-world data with different hardware, jamming strategies, and scenario configurations. We also tested the performance of \sol\ depending on various parameters, such as \ac{SNR} of the communication link, the distance from the jammer and the transmitter, the size of the training set, the number of samples acquired from the channel, the jammer oversampling ratio, and the jamming strategies. Our experimental assessment demonstrates, through an extensive collection of results, the superiority of our solution compared to the current state-of-the-art across all the analyzed configuration parameters.
In general, our solution contributes to taking a step further toward the safe and secure integration of drones into daily life. As part of our future work, we plan to investigate further the effectiveness of \sol\, e.g., when applied for outdoor applications.

\section*{Acknowledgements}

This work has been partially supported by the INTERSECT project, Grant No. NWA.1162.18.301, funded by the Netherlands Organization for Scientific Research (NWO). Any opinions, findings, conclusions, or recommendations expressed in this work are those of the author(s) and do not necessarily reflect the views of NWO. Moreover, this publication was made possible by the NPRP12C-0814-190012-SP165 awards from the Qatar National Research Fund (a member of Qatar Foundation).

\bibliographystyle{IEEEtran}
\balance
\bibliography{refs}

\end{document}